\documentclass[12pt]{iopart} 

\usepackage{graphicx} \usepackage{bm}

\begin{document}

\title[{\it Ab initio} Random Structure Searching]{{\it Ab initio}
  Random Structure Searching}

\author{Chris J Pickard$^{1}$ and R J Needs$^2$}

\address{$^1$\ Department of Physics and Astronomy, University College
  London,\\ Gower St, London WC1E 6BT, United Kingdom}

\address{$^2$\ Theory of Condensed Matter Group, Cavendish Laboratory,
  J J Thomson Avenue, Cambridge CB3 0HE, United Kingdom}

\begin{abstract}
  It is essential to know the arrangement of the atoms in a material
  in order to compute and understand its properties.  Searching for
  stable structures of materials using first-principles electronic
  structure methods, such as density functional theory (DFT), is a
  rapidly growing field.  Here we describe our simple, elegant and
  powerful approach to searching for structures with DFT which we call
  \textit{ab initio} random structure searching (AIRSS).  Applications
  to discovering structures of solids, point defects, surfaces, and
  clusters are reviewed.  New results for iron clusters on graphene,
  silicon clusters, polymeric nitrogen, hydrogen-rich lithium
  hydrides, and boron are presented.
\end{abstract}
 
\submitto{\JPCM}

\maketitle

\section{Introduction \label{sec:introduction}}

Finding the most stable (lowest in energy or free energy) structure of
a large assembly of atoms is a very difficult problem.  The number of
minima in the potential energy surface (PES) of a large system
increases exponentially with the number of atoms.  Finding the global
minimum energy structure with certainty presumably involves visiting
every local minimum and consequently the computational cost also
increases exponentially with the number of atoms.  This effectively
prohibits an exact solution for large systems.  Although the problem
of structure prediction remains very difficult, steady progress has
been made over the years.  Advances in computing power, methods for
calculating accurate energies of assemblies of atoms, and progress in
searching methodologies has led to numerous successful predictions.

Predicting structure is important for a number of reasons.  Structure
prediction is relevant to all areas of science in which one would like
to know the relative positions of atoms.  Computational searching can
be much easier and cheaper than experiments since a range of systems
can quickly be searched, often obtaining interesting results and
sometimes discovering promising new materials.  The low-energy
metastable minima are also interesting as they can be accessed at
finite temperatures, or under pressure.  Structures may also be
trapped in metastable minima during growth or processing.
Computational searches can augment experimental studies when the data
is of poor quality or incomplete.  For example, powder diffraction
data may be insufficient for a complete structural determination but
may suffice to yield information such as the dimensions of the unit
cell and an indication of its likely space group.  The experimental
data can then be used as constraints in a structural search.  The
positions of hydrogen atoms within a crystal cannot easily be
determined from x-ray diffraction data, and here one can use the
positions of the heavier atoms and the dimensions of the unit cell as
constraints.  Computational searches can also be used to investigate
materials under conditions which cannot currently be accessed
experimentally, for example, the pressures within the deep interiors
of massive planets.  Perhaps the most exciting possibility is the
discovery of new materials in the computer which can be synthesised
and have useful applications.

We have used our searching strategy, AIRSS, to predict stable and
metastable structures of crystals and clusters and the atomic
positions at point defects in solids, and we are beginning
applications to surfaces and interfaces.  Only fully quantum
mechanical calculations suffice to deliver the required level of
accuracy because of the wide range of inter-atomic bonding that may be
encountered during the searches.  We calculate the energetics using
first-principles density-functional-theory (DFT) methods
\cite{Hohenberg_1964,Kohn_1966,Payne_1992} which offer a high-level
description of the electronic structure at a cost which is affordable
for the many thousands of structures which must be considered in the
course of a reliable search.

There is a rich literature on computational searching for structures.
It is not our purpose here to review the entire field, although in
\ref{sec:other_searching_methods} we briefly summarise other
approaches to structure searching and give references to the
literature.  In this article we describe our preferred approach in
detail, illustrating the discussion with a variety of examples.

\section{Potential energy surfaces and the global searching
  problem \label{sec:global_searching}}

The exponential increase of the number of local minima with system
size was derived and discussed by Stillinger \cite{Stillinger_1999}.
The basic idea can be gleaned from the following simple argument.
Suppose that a large system of $N$ atoms can be divided into $M$
equivalent subsystems, each of $N/M$ atoms.  If the subsystems are
large enough they will have independent stable configurations.  The
total number of locally stable configurations of the system $n_s$
therefore satisfies
\begin{eqnarray}
\label{eq:number_of_stable_minima_1}
n_s(N) = n_s^M(N/M)  \;.
\end{eqnarray}
The solution to equation (\ref{eq:number_of_stable_minima_1}) is
\begin{eqnarray}
\label{eq:number_of_stable_minima_2}
n_s(N) = e^{\alpha N} \;,
\end{eqnarray}
where $\alpha$ is a constant.  Computational studies of Lennard-Jones
(LJ) clusters support the exponential dependence
\cite{Hoare_1976,Tsai_1993}.

The exponential increase in the number of local minima suggests that
it will be very difficult to devise a reliable approach for finding
the global minimum energy state of a large system.  Perhaps clever
methods can be found for eliminating the exponential scaling?
Although it is not currently possible to give a definitive answer to
this question, the prospects appear bleak.  Determining the global
minimum of a PES is classed as an NP-hard (non-deterministic
polynomial-time hard) problem.  These are problems for which it is
widely suspected (but not proven) that it is impossible to find an
algorithm which works without fail in polynomial time.  Reducing the
strength of the exponential scaling (i.e., reducing the value of
$\alpha$ in equation (\ref{eq:number_of_stable_minima_2})) is a more
realistic goal, but theory also provides us with a warning about this.
Wolpert and Macready have proved a ``no free lunch theorem'' for
searching and optimisation which shows that all algorithms that search
for the global minimum of an energy function perform exactly the same
when averaged over all possible energy functions \cite{Wolpert_1997}.
The implication is that it may be extremely difficult or even
impossible to find a smart algorithm which works well in all
circumstances.

We are interested in the energy functions which represent the PES of
assemblies of atoms, and these form only a very small subset of all
possible energy functions.  Much of the PES of a reasonably large
assembly of atoms corresponds to very high energy structures in which
some atoms are much closer than an equilibrium bond length.  This can
readily be verified by calculating the energies of an ensemble of
``random'' structures, each formed by placing atoms at random
positions within a box whose size gives a physically reasonable
density.  The average energy will be far higher than even the highest
energy local minimum because of the strong short-range atomic
repulsion.  Other parts of the PES will correspond to fragmented
structures.  These may contain interesting energy minima, but if we
are only interested in fully connected structures we can disregard
them.

A basin of attraction of a PES is defined as the set of points for
which downhill relaxation leads to the same energy minimum.  A PES can
therefore be divided into basins of attraction.  Some rather general
features of the PES of an assembly of atoms and its basins of
attraction are known:\\
\textbf{(i)} The substantial fraction of the PES in which some atoms
are very close together contains almost no minima.\\
\textbf{(ii)} The basins are normally arranged such that if one moves
from a basin to a neighbour it is more likely that the neighbour will
have a lower energy minimum if the barrier between the basins is
small.  This is a consequence of the relative smoothness of the PES at
low energies and is related to the Bell-Evans-Polanyi principle which
states that highly exothermic chemical reactions have
low activation energies \cite{jensen_1999}.\\
\textbf{(iii)} Another implication of the Bell-Evans-Polanyi principle
is that low energy basins are expected to occur near other low energy
basins.  Of course low energy basins can occur in widely separated
``clumps'', which are normally referred to as ``funnels''.\\
\textbf{(iv)} The probability distribution of the energies of the
local minima of a PES is close to Gaussian for large systems, as would
apply for the model which leads to equation
(\ref{eq:number_of_stable_minima_2}).\\
\textbf{(v)} Various studies have shown that basins with lower energy
minima tend to have larger hyper-volumes in the ``structure space''
than higher energy minima \cite{Doye_1998,Doye_2005b}.\\
\textbf{(vi)} The probability distribution of the hyper-volumes of the
basins appear to decrease as a power law in the minimum energy of a
basin \cite{Massen_2007b}.  It seems that the power law behaviour must
derive from some type of order in the arrangement of basins of
different sizes, with smaller basins filling the gaps between larger
ones \cite{Massen_2007a}.  The power law distribution does not occur
in a simple model PES formed by arranging Gaussians of random width
\cite{Massen_2007a}.\\
\textbf{(vii)} Both very-low (and very-high) energy minima tend to
correspond to symmetrical structures.  The tendency of low-energy
minima to be symmetrical is supported by the ubiquity of crystals and
is related to Pauling's ``rule of parsimony'' which states that ``The
number of essentially different kinds of constituents in a crystal
tends to be small'' \cite{Pauling_1929}.  The symmetry of both
very-low and high energy minima is also supported by calculations
\cite{Wales_1998,Wales_1998_erratum}.\\

\textbf{(viii)} It has been observed that some space group symmetries
are much more common than others in crystals formed from small organic
molecules \cite{Nowacki_1943,Mighell_1983,Donohue_1985}.  Inorganic
systems show different space group frequencies
\cite{Mackay_1967,Urusov_2009}.\\
\textbf{(ix)} As well as general features of the PES of assemblies of
atoms, there are particular features which arise from chemical
considerations.  In fact we normally know a great deal about the
chemistry of the systems we study.  We often know which atomic types
prefer to bond to one another and the approximate lengths of the
bonds, and the likely coordination numbers of the atoms.

\section{Random Structure
  Searching \label{sec:random_structure_searching}}

If nothing is known about the likely low-energy structures it is
reasonable to start searching by relaxing random structures, which
gives the widest coverage of the PES and an unbiased sampling.  The
notion of ``random structures'' is explored in Section
\ref{subsec:generating_random_structures}, and it will turn out that
we must impose limits on the initial structures for reasons of
efficiency, so that our ``random structures'' might better be
described as ``random sensible structures''.  Using random sensible
structures is a useful approach which we have used successfully in
several of the applications described in Section
\ref{sec:airss_calculations}.  Given \textbf{(iv)} (that the
distribution of the energies of the local minima is approximately
Gaussian), it may seem surprising that random structure searching
works at all.  However, features \textbf{(i)} (there are almost no
minima at high energies), \textbf{(v)} (low-energy minima have large
hyper-volumes), and \textbf{(vi)} (the distribution of
the hyper-volumes of the basins follow a power law), act in favour of
the searcher.

These features imply that even random sampling has a good chance of
finding low energy basins and that the wide coverage of the PES gives
a chance of sampling the different ``funnels'' mentioned in
\textbf{(iii)}.  We exploit features \textbf{(vii)} and
\textbf{(viii)} by imposing symmetry constraints as explained in
Section \ref{subsec:Imposing symmetry}.  We make use of the proximity
of low-energy basins of \textbf{(iii)} by ``shaking'' structures so
that they fall into nearby minima, see Section \ref{subsec:shaking}.
Following \textbf{(ix)}, we also make extensive use of chemical
understanding of the system, as described in Section
\ref{subsec:Imposing chemical ideas}.

Our approach is very simple as it requires very few parameters and is
very easy to implement.  The biases are largely controllable,
understandable, and based on sound principles.  The searches run very
efficiently on modern parallel computers.  Our experience with the
primitive method has been that we can perform highly reliable searches
for the global minimum with up to at least 12 atoms (of one or two
species) and often more.  When imposing constraints we can search
successfully on much larger systems.  Information from experiments,
and chemical and structural information for the system in question or
similar systems, and information generated by previous searches are
combined to help design searches.  The most successful approaches to
searching are those which make the best use of the available
information to bias the search towards finding the desired structures.

Our searches find many local minima, particularly if constraints are
not imposed.  As mentioned in Section \ref{sec:introduction}, it is
not only the ground state structure which is of interest, higher
energy structures can also be important.  For example, technologies
such as molecular beam epitaxy (MBE) and Metal-Organic Chemical Vapour
Deposition (MOCVD) allow controlled epitaxial growth of materials,
which can result in structures far from equilibrium.  Structure
searching allows the discovery of many possible stable and metastable
materials, which can then be ranked according to any property of
interest such as the band gap or bulk modulus.  In our work we
emphasise the possibilities of discovering low-energy structures
rather than designing structures with particular properties, because
only low-energy structures can normally be synthesised.

Random structure searching also teaches us chemistry.  For example, we
threw hydrogen atoms (H) and oxygen atoms (O) in the ratio 2:1 into a
box and relaxed, finding the most stable structures to consist of
H$_2$O molecules.  Of course we expected this but, studying the
higher-energy structures, we found other low-energy small molecules
composed of H and O atoms \cite{Pickard_2007_water}.

\subsection{Generating random structures \label{subsec:generating_random_structures}}

\begin{figure}[ht!]
\centering
\includegraphics[width=0.75\textwidth]{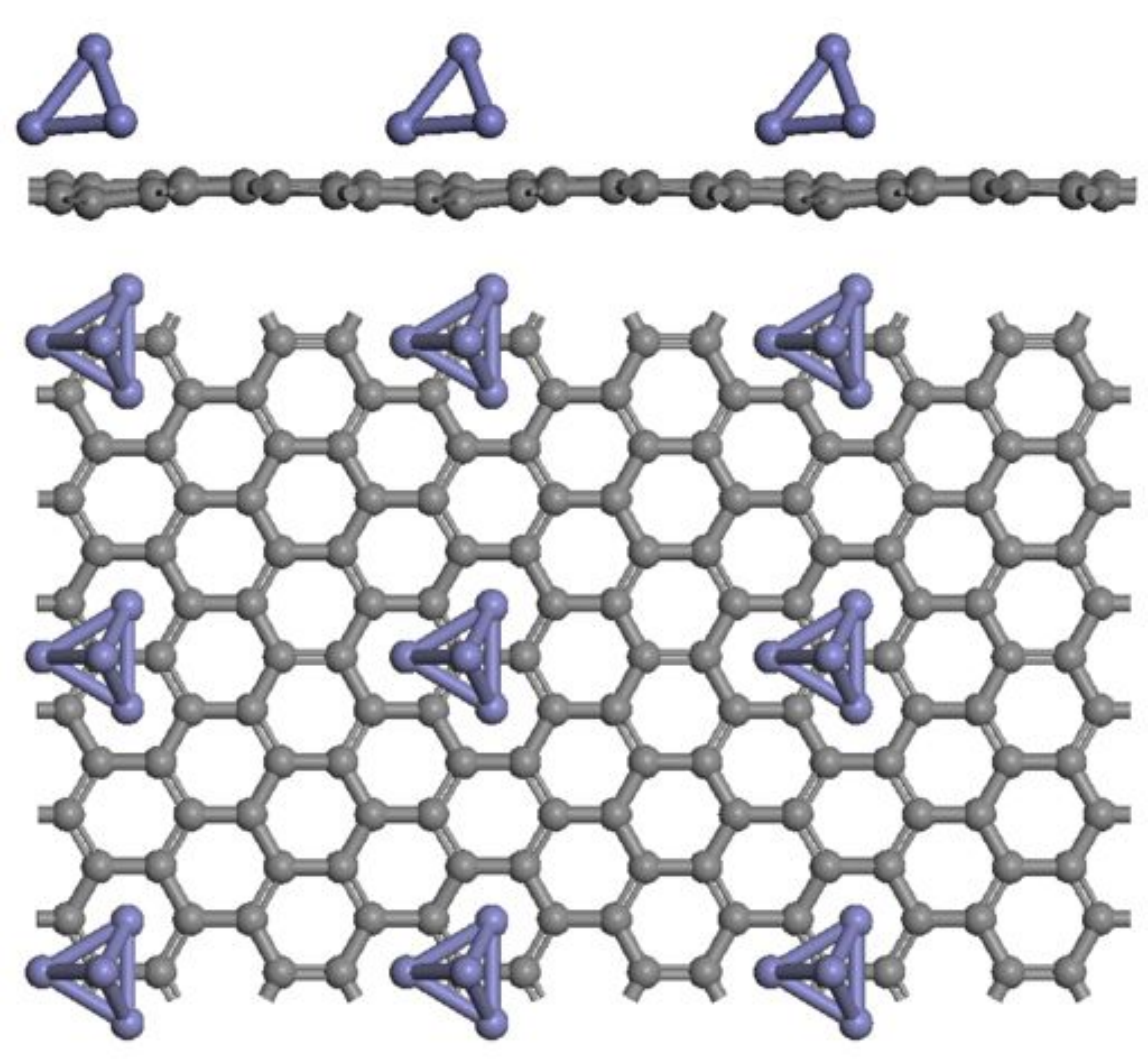}
\caption[]{Iron clusters on graphene.  A (non-magnetic) four-atom iron
  cluster of random shape was placed at a random position on top of a
  graphene sheet represented by a 24-atom supercell, and all the
  atomic positions were relaxed.  The lowest energy structure obtained
  after relaxing 69 structures is shown.  Note the distorted
  tetrahedron of the iron cluster and how well the iron cluster
  matches the graphene lattice.}
\label{fig:fe_on_graphene}    
\end{figure}

What do we mean by the term ``random structure''?  The atomic
arrangements in real materials are not at all random because the
diameters of atoms and the bond lengths between them lie within a
rather small range of roughly 0.75 to 3 \AA.  An assembly of atoms
therefore has a ``natural volume'' which is proportional to the number
of atoms present but only rather weakly dependent on the identities of
the atoms and the external conditions.  We start searches from
fully-connected structures because separate fragments do not ``see''
each other and are unlikely to join up during relaxation.  We adopt
different procedures for generating initial structures for bulk
solids, clusters and point defects in solids.  Procedures can easily
be devised for other purposes such as finding surface or interface
structures, see figure \ref{fig:fe_on_graphene}.

\subsection{{\bf Periodic solids}} 

A random set of unit cell lengths $(a,b,c)$ and angles
$(\alpha,\beta,\gamma)$ is chosen and the cell volume is renormalised
to a random value within $\pm$50\% (or thereabouts) of a chosen mean
volume.  An appropriate mean volume can be determined from known
structures composed of the same atoms, by adding up atomic volumes, or
by relaxing a few ``handmade'' structures.  The results are not very
sensitive to the mean volume and range chosen.  It turns out that a
unit cell with very large or small angles can be transformed into an
entirely equivalent unit cell with angles in the range
60$^{\circ}$--120$^{\circ}$.  The more compact transformed cells are
helpful for choosing efficient grids for Brillouin zone integrations
and in visualising structures.  We transform to more compact cells
whenever possible.

\subsection{{\bf Clusters}}

\begin{figure}[ht!]
\centering
\includegraphics[width=0.75\textwidth]{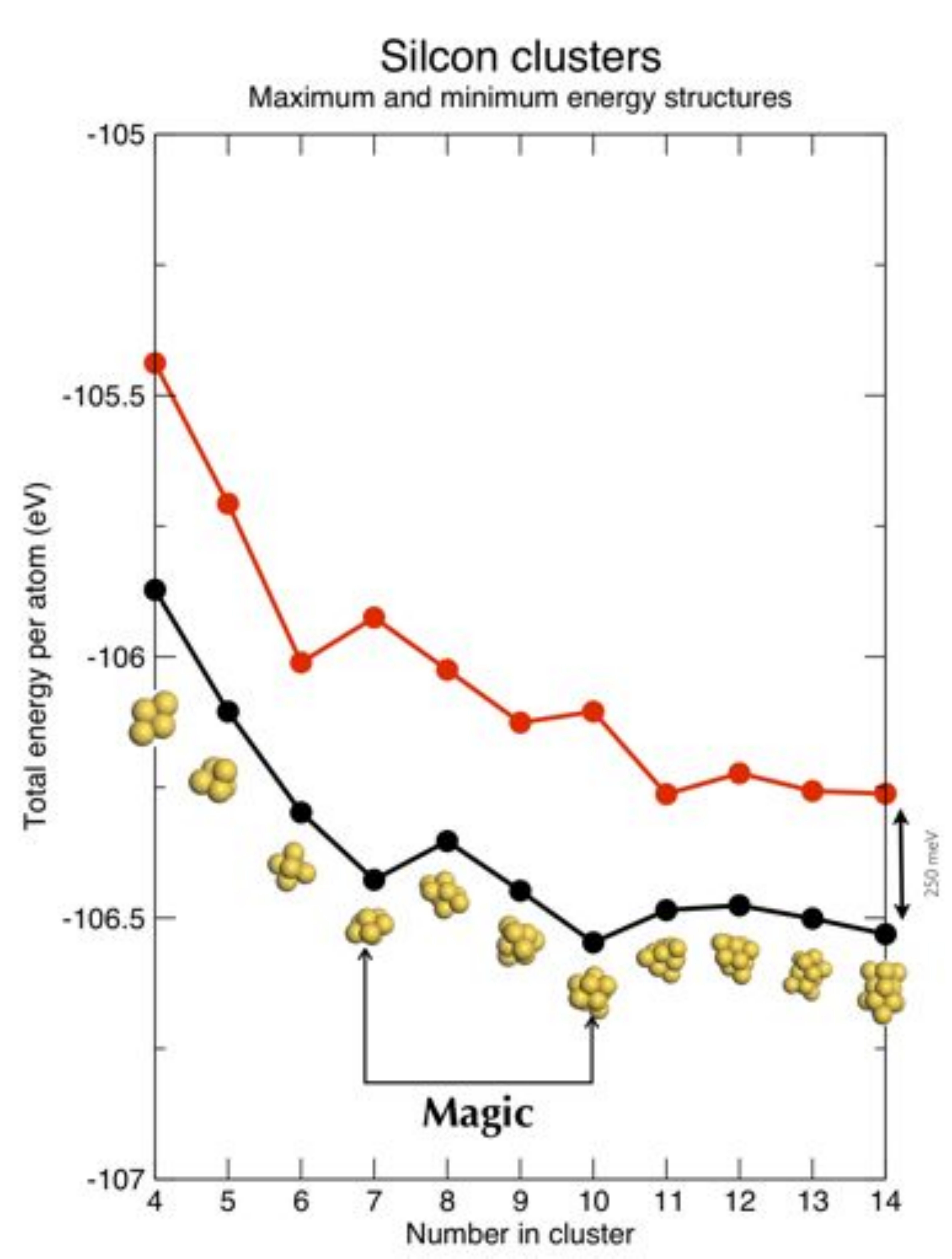}
\caption[]{Silicon clusters were generated by placing atoms randomly
  within a small box inside a large unit cell and relaxing within DFT.
  The algorithm generated the same lowest-energy structures obtained
  in previous DFT studies \cite{Cambridge_Cluster_Database}, including
  the two ``magic'' number clusters with seven and ten atoms.  We also
  found many local minima.  The highest-energy minimum for each
  cluster size is only about 0.25 eV per atom higher in energy than
  the minimum energy structure.}
\label{fig:Cluster}    
\end{figure}

To generate initial structures for clusters we choose a box/sphere of
a reasonable size to enclose the cluster and insert the atoms at
random, as in a calculation for a periodic solid.  We then place the
box/sphere inside a considerably larger unit cell and impose periodic
boundary conditions.  An example of searching in clusters is described
in figure \ref{fig:Cluster}.

\subsection{{\bf Point defects}}

We start from a supercell of the perfect host crystal.  In our work on
defects in diamond-structure semiconductors (see Section
\ref{sec:airss_calculations}) we have mostly used 32-atom supercells,
although some defects may require larger cells.  We remove a few
neighbouring atoms from the crystal to make a ``hole'', into which we
place at random the desired host and impurity atoms.

\subsection{{\bf Keeping atoms/molecules apart}}

Random structures may contain atoms which are very close together.
Such occurrences are often harmless as the forces on the atoms are
very large and they quickly move apart under relaxation.  We have,
however, sometimes encountered problems when transition metal atoms
are nearly on top of one another which can make it very difficult to
achieve self-consistency so that accurate forces cannot be obtained.
A related problem occurs in searching for the structures of molecular
crystals, where starting from randomly placed molecules can lead to
unwanted chemical reactions.  These difficulties can be avoided by
rejecting starting structures in which atoms or molecules are too
close.  For very large systems the fraction of structures rejected
will approach unity and a more efficient procedure should be used in
which atoms or molecules are ``nudged'' apart.

\section{Biasing the searches  \label{sec:biasing_searches}}
 
\subsection{{\bf Choosing stoichiometries}}

Does element A react with element B to form the compound AB, or
perhaps A$_2$B, or A$_2$B$_3$ etc., or is the compound A$_2$B$_3$
unstable to the formation of A$_2$B + 2B, or 2AB + B ?  These
questions can be answered by determining the energies of the most
stable structures of each compound, which allows the thermodynamically
most stable state of a mixture of A and B to be determined.  This
problem involves searching a larger structure space than is
required for determining the most stable structure of a particular
stoichiometry, but it can be tackled by carrying out structural
searches for a range of stoichiometries.  Searching with a particular
stoichiometry may give hints about more stable stoichiometries as
phase separation can occur within the unit cell.  We have often
noticed such behaviour although the limited size of the cells means
that calculations with other stoichiometries and cell sizes may be
necessary to unambiguously identify phase separation.  An example of
searches over different stoichiometries is described in figure
\ref{fig:Li+H}.  The first source of bias in studying a system is
therefore the choice of stoichiometries.

\subsection{{\bf Choosing the number of units}}

When searching for crystalline phases of a given stoichiometry one
does not \textit{a priori} know how many formula units the primitive
unit cell contains, and one should perform searches with different
numbers of units.  Searching using ``usual'' numbers of formula units,
such as 2, 4, 6, and 8, will normally be an effective way to bias the
search. However, it will preclude unexpected results, for example the
11 and 21 atom host-guest phases of aluminium (discussed in Section
\ref{subsec:Aluminium}).  We are fighting a computational cost that
grows rapidly with system size and performing nearly exhaustive
searches with more formula units rapidly becomes impracticable.
Random structures are a perfectly reasonable starting point if one has
no knowledge of the likely structures, but with a little thought one
can often greatly improve the efficiency of the search by biasing it
towards finding low energy structures.  This makes it possible to
perform more comprehensive searches with larger numbers of atoms.

\subsection{{\bf Imposing chemical ideas} \label{subsec:Imposing chemical ideas}}

\begin{figure}[ht!]
\centering
\includegraphics[width=0.75\textwidth]{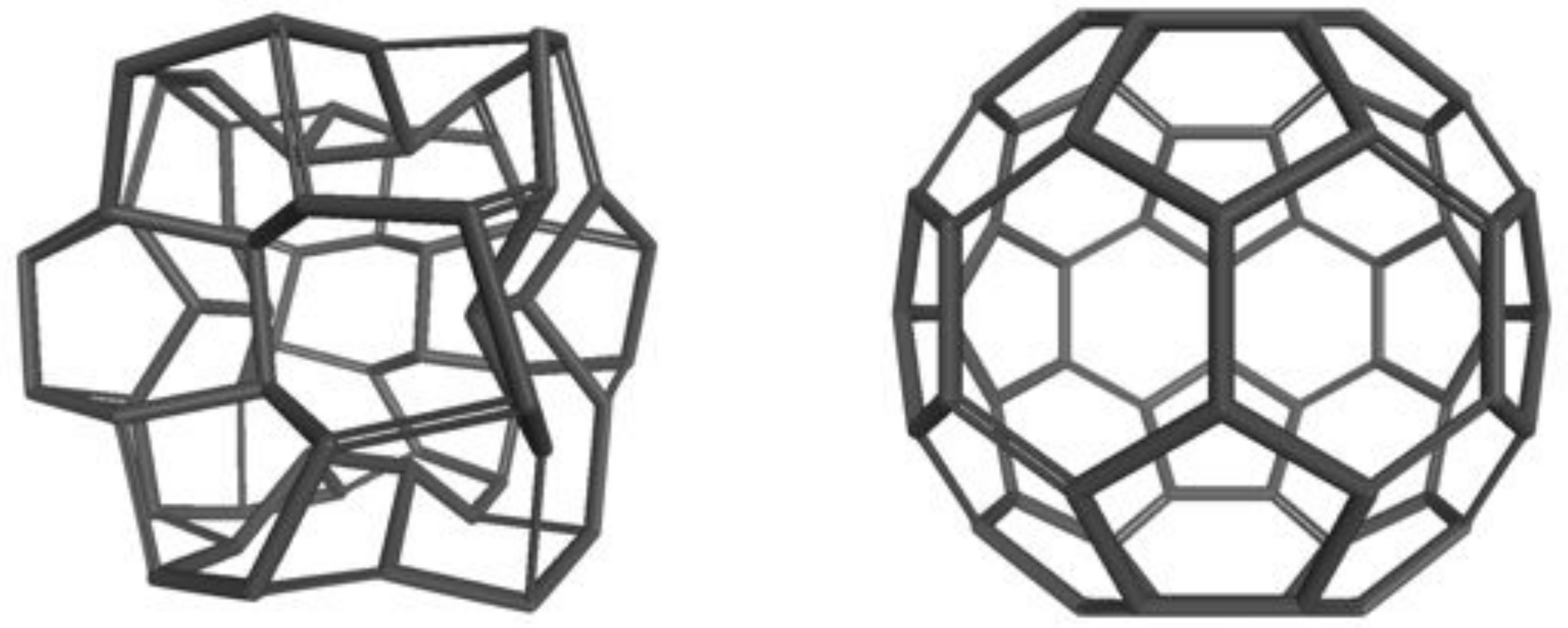}
\caption[]{Left: A structure built by placing carbon atoms randomly
  within a small sub-box, subject to symmetry constraints.  Random
  structures were generated and then screened to determine whether the
  atoms were three-fold coordinated.  If not, the structure was
  rejected and another one was generated.  Right: relaxation of this
  structure within DFT gave the well-known C$_{60}$ ``buckyball''.}
\label{fig:C60}    
\end{figure}

Extensive knowledge of the chemistry of a system is often available,
even if we know little about the actual structures which are favoured.
Under these circumstances one can use chemical ideas to bias the
searching.  We already mentioned the idea of choosing initial
structures composed of molecular units, and other examples of imposing
chemical ideas are discussed in Section\
\ref{sec:random_structure_searching}.  Even if the system is
non-molecular it is often possible to use chemical units to increase
the efficiency of the search.  For example, if one is interested in
structures of silicon dioxide one can make initial structures from
O--Si--O units.  This has the effect of making the densities of the Si
and O atoms much more uniform than a random structure, which becomes
increasingly important for larger system sizes, and biasing towards
the correct bonding.  Another important chemical idea is that of
coordination number.  For example, we can generate initial structures
of carbon with $sp^2$ bonding by creating random structures and
rejecting all those which are not 3-fold coordinated, as illustrated
in figure \ref{fig:C60}.

\subsection{{\bf Imposing symmetry} \label{subsec:Imposing symmetry}}

As noted in Section \ref{sec:global_searching}, minima with very low
or very high energies tend to correspond to symmetrical structures.
Imposing a degree of symmetry on the initial structures and
maintaining it during relaxation therefore eliminates a large amount
of the PES while (hopefully) still allowing the global minimum energy
structure to be found.  We implement this strategy by searching
randomly over all space groups with $N_s$ symmetry operations.  Such a
search also allows structures to relax into space groups which are
super-groups of those with $N_s$ symmetry operations.  Symmetry
constraints have often been used in searching for crystalline
polymorphs composed of small molecules such as the drug molecules
developed within the pharmaceutical industry \cite{price_2009}.

\subsection{{\bf Using experimental data}}

\begin{figure}[ht!]
\centering
\includegraphics[width=0.75\textwidth]{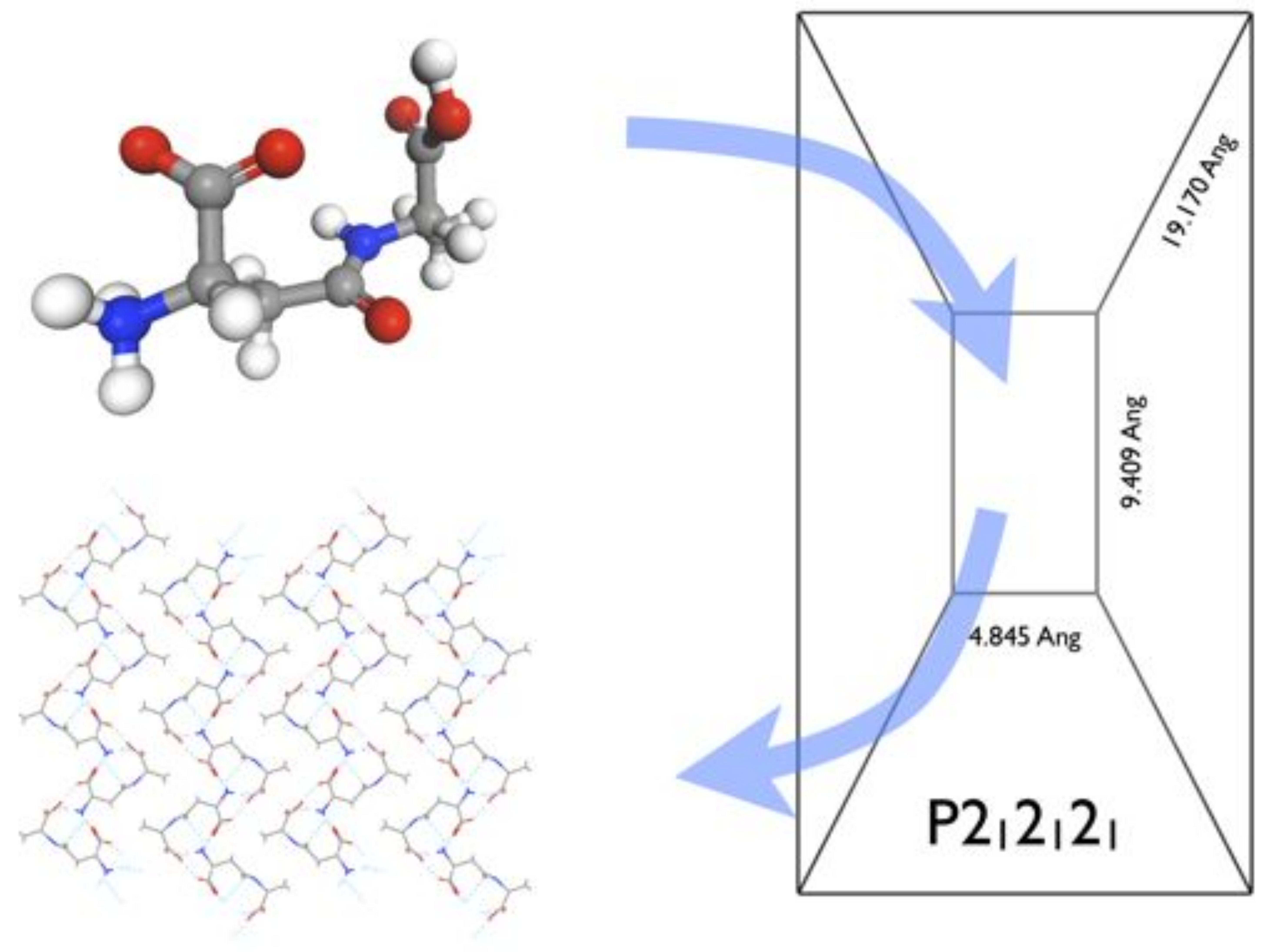}
\caption[]{The crystal structure of the beta-L-aspartyl-L-alanine
  dipeptide is known experimentally.  In this test we made structures
  from the experimental unit cell shown on the right and the
  $P2_12_12_1$ space group of the crystal and the structure of the
  beta-L-aspartyl-L-alanine molecule (top left).  Carbon atoms are
  shown in grey, oxygen in red, nitrogen in blue and hydrogen in
  white.  A single molecule was placed randomly in the unit cell and
  the positions of the other three molecules were determined by the
  space group symmetry.  The structure was rejected if two molecules
  overlapped and a new one was generated.  Each non-overlapping
  structure was relaxed within DFT while maintaining the size and
  shape of the unit cell and the $P2_12_12_1$ symmetry.  The correct
  molecular packing (bottom left) was found after relaxing 18
  structures.}
\label{fig:experiment_dipeptide}    
\end{figure}

We already mentioned in Sections \ref{sec:introduction} and
\ref{sec:global_searching} the possibility of using experimental data
to bias a search.  It may turn out that a powder diffraction spectrum
is obtained with quite a few well defined peaks which, however, are
insufficient for a full structural determination.  In such cases it is
often possible to determine the dimensions of the unit cell and
perhaps an indication of the most likely space groups from the data.
Such information is extremely useful when performing a structural
search, and an example of this type of constrained search is described
in Section \ref{Ammonia monohydrate} for a high-pressure phase of
ammonia monohydrate, and a test calculation for a dipeptide is
illustrated in figure \ref{fig:experiment_dipeptide}.  Knowledge of
the different space group frequencies, which we mentioned in Section
\ref{sec:global_searching}, could also be used to bias searches.

\subsection{{\bf Shaking} \label{subsec:shaking}}

In Section \ref{sec:global_searching} we encountered the idea that low
energy basins may be clustered together.  This motivates the
``shake'', a random displacement of the atoms and, if appropriate, a
random adjustment of the unit cell.  Atomic displacements of a large
fraction of a bond length have a reasonable chance of pushing the
system into a nearby basin of attraction.  We have also used shaking
to look for distortions of structures into doubled (or larger) unit
cells \cite{LePage_2006}.  The shake is the same as a step in the
basin hopping algorithm \cite{Li_1987,Wales_1999,Wales_book_2003} (see
also, \ref{sec:other_searching_methods}), although we have used it
only with zero temperature and after considerable searching has
already located low-energy structures.  Shaking can also be integrated
into searching strategies as in ``relax and shake'' (RASH), see
Section \ref{sec:lessons_searching_simple_potentials}.

A related idea is to calculate the harmonic phonon modes of a
structure.  The phonon modes at zero wave vector of a fully relaxed
structure found from unconstrained random searching must be stable,
and the structure must also be stable against elastic distortions.
The phonon modes at non-zero wave vectors may, however, be unstable,
so that the energy can be reduced by a distortion in a larger unit
cell.  Calculating the second derivatives of the energy to obtain the
phonon frequencies and displacement patterns is expensive and we only
perform such calculations on a few structures of interest after
extensive searching.  If unstable phonon modes are found then the
energy-reducing distortions of the corresponding phonon eigenvectors
can be followed to find more stable structures.

\section{Have we found the global minimum?}

The searching is not exhaustive and therefore we cannot be sure that
we have found the global minimum.  One way to gauge the quality of a
search is to look for known ``marker'' structures (if available).  We
happily terminate searches when the same lowest-energy structure has
been found several times.  This criterion is reasonable because we
relax a very wide range of initial structures.  When we apply
constraints to the initial structures and maintain them during the
relaxation we obviously cannot obtain structures which violate the
constraints.  When we apply constraints to the initial structures but
allow free relaxation we are biasing the search, presumably towards
structures which obey the initial constraints, but also perhaps in
ways which we cannot predict.  When we bias a search it is important
to understand as well as possible which parts of the PES are being
excluded or de-emphasised.  This allows the user to assess the
strengths and weaknesses of a search and, if required, to design
further searches.  It is therefore important that the effects of the
``knobs'' of the search (the parameters which can be varied) are as
transparent as possible.  The simplicity of our searching procedures
results in a relatively small number of understandable and useful
knobs.  This makes it easy to decide on appropriate values for any
variable parameters of the search, so that costly trials are not
required to optimise the search procedure.

\section{Some technical aspects of the calculations \label{sec:technical}}

\subsection{{\bf First-principles DFT calculations}}

DFT calculations are much more expensive than empirical potential ones
and the number of structures whose energies may be evaluated is
therefore greatly fewer.  Many first-principles DFT codes are
available, and we use the \textsc{CASTEP} package \cite{ClarkSPHPRP05}
which uses a plane wave basis set, periodic boundary conditions, and
pseudopotentials.  The code returns the total energy of a structure
and the forces on the atoms and stresses on the unit cell.  We use the
forces and stresses to relax structures to the nearest local minimum
in the PES.  The second derivatives of the energy may be calculated by
linear response or finite displacement methods \cite{baroni_2001}, and
although these methods are very useful in checking for unstable
phonons/elastic distortions and in calculating thermal effects in
stable structures, they are far too expensive to be used routinely as
part of the search strategy.

\subsection{{\bf Pseudopotentials}}

Accurate results at very high pressures can be obtained using
pseudopotentials, but they must be constructed with sufficiently small
core radii and with the appropriate electrons treated explicitly.  The
pseudopotentials provided with standard codes may be inadequate at the
high pressures we often work at.  Lithium is an unusually difficult
case.  It is standard to treat all three electrons of lithium
explicitly, but the pseudopotential core radii must still be small
\cite{Pickard_2009_lithium} in high-pressure studies.  We use
ultrasoft pseudopotentials \cite{Vanderbilt90} and find them to be
accurate when the distance between neighbouring atoms is about equal
to or greater than the sum of the core radii of the atoms.  We
recommend that pseudopotentials be tested for each application by
generating them with different core radii and checking that energy
differences are accurate for the shortest inter-atomic distances that
will be encountered.  For some of the applications described in
Section \ref{sec:airss_calculations} we have treated some core and
semi-core states explicitly.  For example, we used pseudopotentials
with 11 electrons treated explicitly for our work on aluminium
\cite{Pickard_2010_aluminium} and 16 electrons for iron
\cite{Pickard_2009_iron}.

\subsection{{\bf k-point sampling}}

We use quite good Brillouin sampling and basis sets when searching
because we find that poor quality calculations can lead to strong
biases.  We have come across modulated phases when searching in metals
which went away when we relaxed them further with denser k-point
sampling.  We use Monkhorst-Pack (MP) meshes of k-points
\cite{monkhorst_1976} which are defined by choosing the smallest MP
mesh for which the smallest separation between k-points is less than
some distance $\Delta k$.  We often use $\Delta k =
2\pi\times$0.07~\AA$^{-1}$ when searching and then perhaps $\Delta k =
2\pi\times$0.03~\AA$^{-1}$ when refining the structures and their
energetics.  We deform the k-point mesh with the changes in the cell
shape and occasionally recalculate the integer parameters of the MP
mesh.

\subsection{{\bf Predicting stability over a range of pressures}}

In our high-pressure studies we search at constant pressure, although
one can just as easily search at constant volume.  A search at
pressure $p_{\rm s}$ may give many different structures.  The
structure with the lowest enthalpy $H(p_{\rm s})$ is the most stable
at $p_{\rm s}$, but different structures may be more stable at another
pressure $p$.  To investigate this we can use the thermodynamic
relation
\begin{eqnarray}
H(p) & \simeq & H(p_{\rm s}) + (p-p_{\rm s}) \left. \frac{dH}{dp}\right|_{p_{\rm s}} 
  + \frac{1}{2}  (p-p_{\rm s})^2 \left. \frac{d^2H}{dp^2}\right|_{p_{\rm s}} \\
  & = & H(p_{\rm s}) + (p-p_{\rm s}) V_{\rm s} 
  - \frac{1}{2} (p-p_{\rm s})^2 \frac{V_{\rm s}}{B_{\rm s}} \;,
\label{eq:enthalpy-volume_1}
\end{eqnarray}
where $V_{\rm s}$ and $B_{\rm s}$ are the volume and bulk modulus at
It may be possible to use the quadratic form of equation
(\ref{eq:enthalpy-volume_1}) with an empirical relationship between
the bulk modulus and volume, but we have not explored this further.
We have found the simple linear approximation 
\begin{eqnarray}
\label{eq:enthalpy-volume_2} H(p) & \simeq & H(p_{\rm s}) + 
(p-p_{\rm s}) V_{\rm s} 
\end{eqnarray}
to be particularly convenient because the quantities required
($H(p_{\rm s})$, $p_{\rm s}$, $V_{\rm s}$) are calculated for each
relaxed structure obtained in a search.  The data can then be used to
estimate the stability regions of the different structures over a wide
range of pressures, which gives the approach a ``far sightedness''.

\begin{figure}[ht!]
\centering
\includegraphics[width=0.7\textwidth]{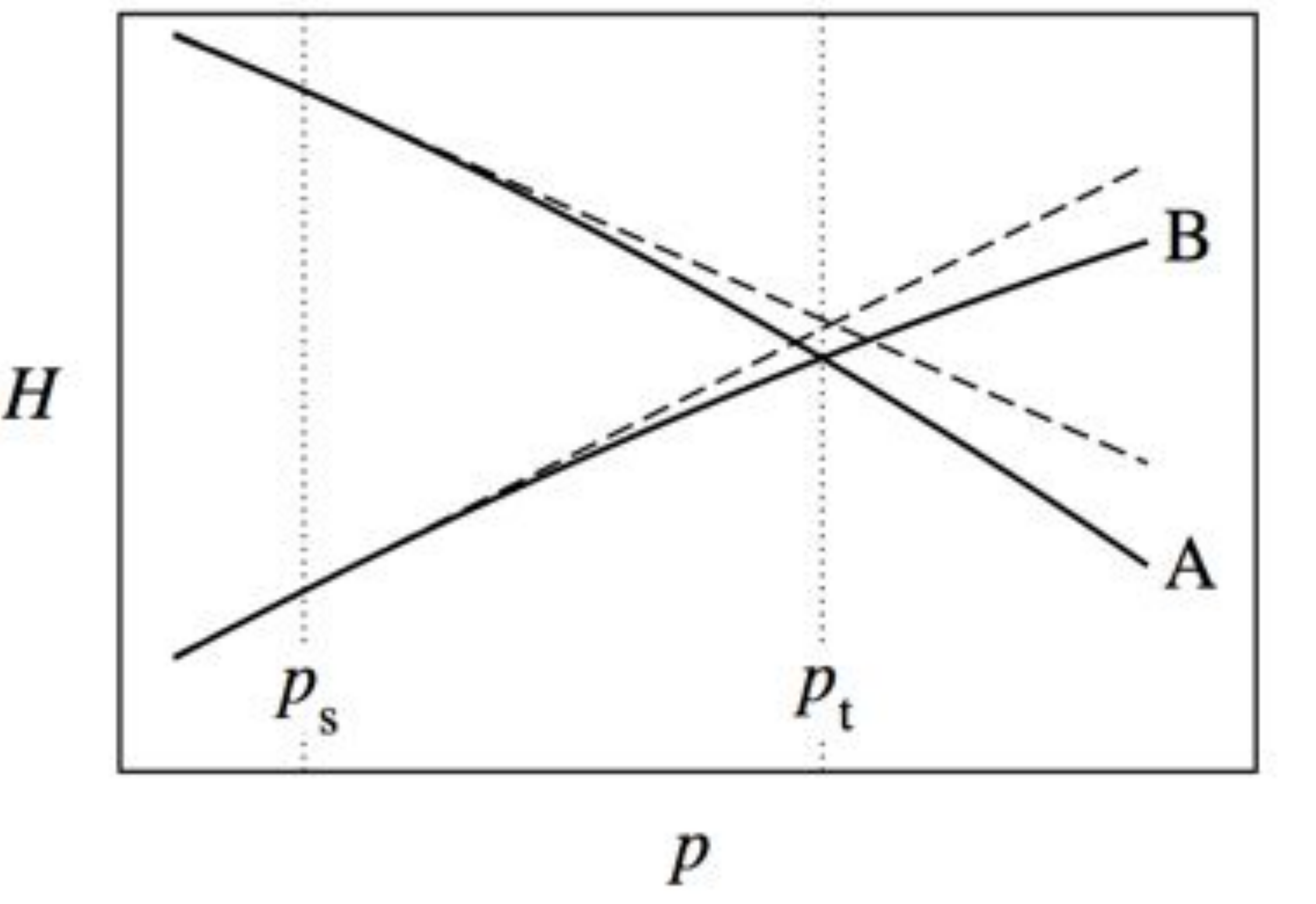}
\caption[]{The solid lines show the variation of the enthalpy $H$ with
  pressure $p$ for two phases A and B.  The search is performed at
  pressure $p_{\rm s}$ and the phase transition from B to A occurs at
  pressure $p_{\rm t}$.  The enthalpies predicted by the linear
  approximation of equation (\ref{eq:enthalpy-volume_2}) are shown as
  dashed lines.  The linear approximation gives a transition pressure
  close to $p_{\rm t}$.}
\label{fig:enthalpy-pressure}
\end{figure}

\begin{figure}[ht!]
\centering
\includegraphics[width=0.85\textwidth]{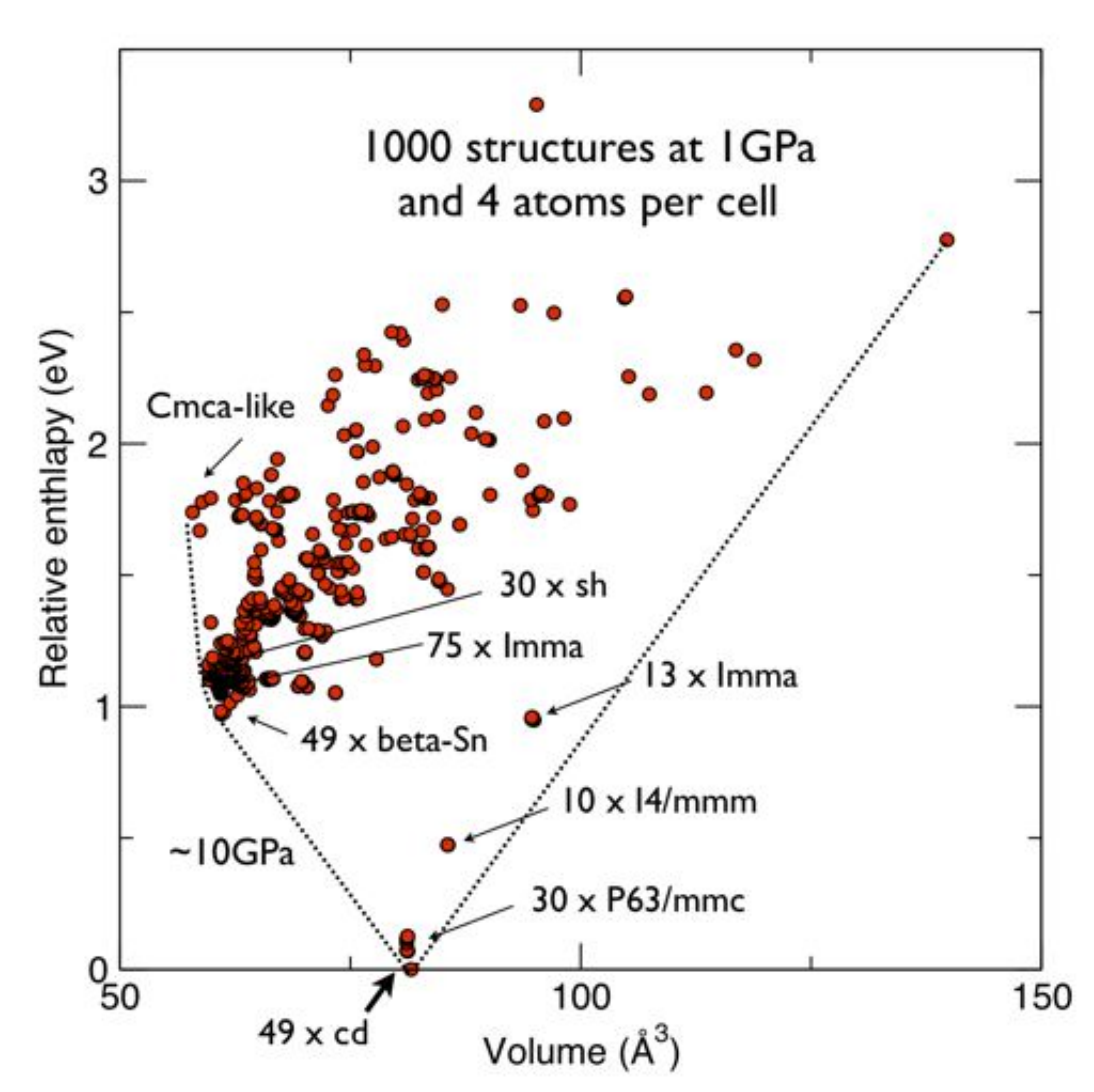}
\caption[]{Scatter plot of the relative enthalpies against volume for
  a search at $p_{\rm s}$ = 1~GPa with 4 silicon atoms per cell.  Both
  the enthalpies and volumes are given for 4 atoms.  The diamond
  structure (cd) is the most stable at this pressure and was found 49
  times from a total of 1000 relaxed structures.  The positions of the
  observed high-pressure phases \cite{mujica_review_03}, beta-Sn,
  $Imma$, sh (simple hexagonal) and $Cmca$-like, are also indicated on
  the figure.  The hexagonal-close-packed (hcp) and face-centred-cubic
  (fcc) phases which are observed in experiments at pressures beyond
  the $Cmca$ phase \cite{mujica_review_03} were not found in the
  searches and we presume they are mechanically unstable at 1~GPa.
  Equation (\ref{eq:enthalpy-volume_2}) shows that the stable phases
  can be found by drawing lines underneath the data points as shown in
  the figure.  The stable phases at pressures greater than 1~GPa can
  then be read off the figure as those through which the dotted lines
  pass, and it can be seen that these are the experimentally observed
  ones at positive pressures where the dotted line has a negative
  slope.  The slope of the line joining the cd and beta-Sn phases
  corresponds to a pressure of about 10~GPa, which is similar to the
  coexistence pressure \cite{mujica_review_03}. The phases above the
  dotted lines are not the most stable at any pressure.  The
  $P6_3/mmc$ phase differs from cd only in the stacking of layers.}
\label{fig:enthalpy_volume_scatter}
\end{figure}

The application of equation (\ref{eq:enthalpy-volume_2}) to two
structures, A and B, found at $p_{\rm s}$ is illustrated in figure
\ref{fig:enthalpy-pressure}.  Equation (\ref{eq:enthalpy-volume_2})
tells us that if $V_{\rm s}^{\rm A} < V_{\rm s}^{\rm B}$ then
structure A will become more favourable with respect to structure B at
$p > p_{\rm s}$ and less favourable for $p < p_{\rm s}$.  If B is more
stable than A at $p_{\rm s}$ a phase transition from B to A could
occur at some $p > p_{\rm s}$.  Equations (\ref{eq:enthalpy-volume_1})
and (\ref{eq:enthalpy-volume_2}) can be applied to results obtained in
both constant volume and constant pressure calculations, although we
often use a simple scatter-diagram representation in our constant
pressure calculations, as explained in figure
\ref{fig:enthalpy_volume_scatter}.

\section{Lessons from searching with simple
  potentials \label{sec:lessons_searching_simple_potentials}}

\begin{figure}[ht!]
\centering
\includegraphics[width=0.85\textwidth]{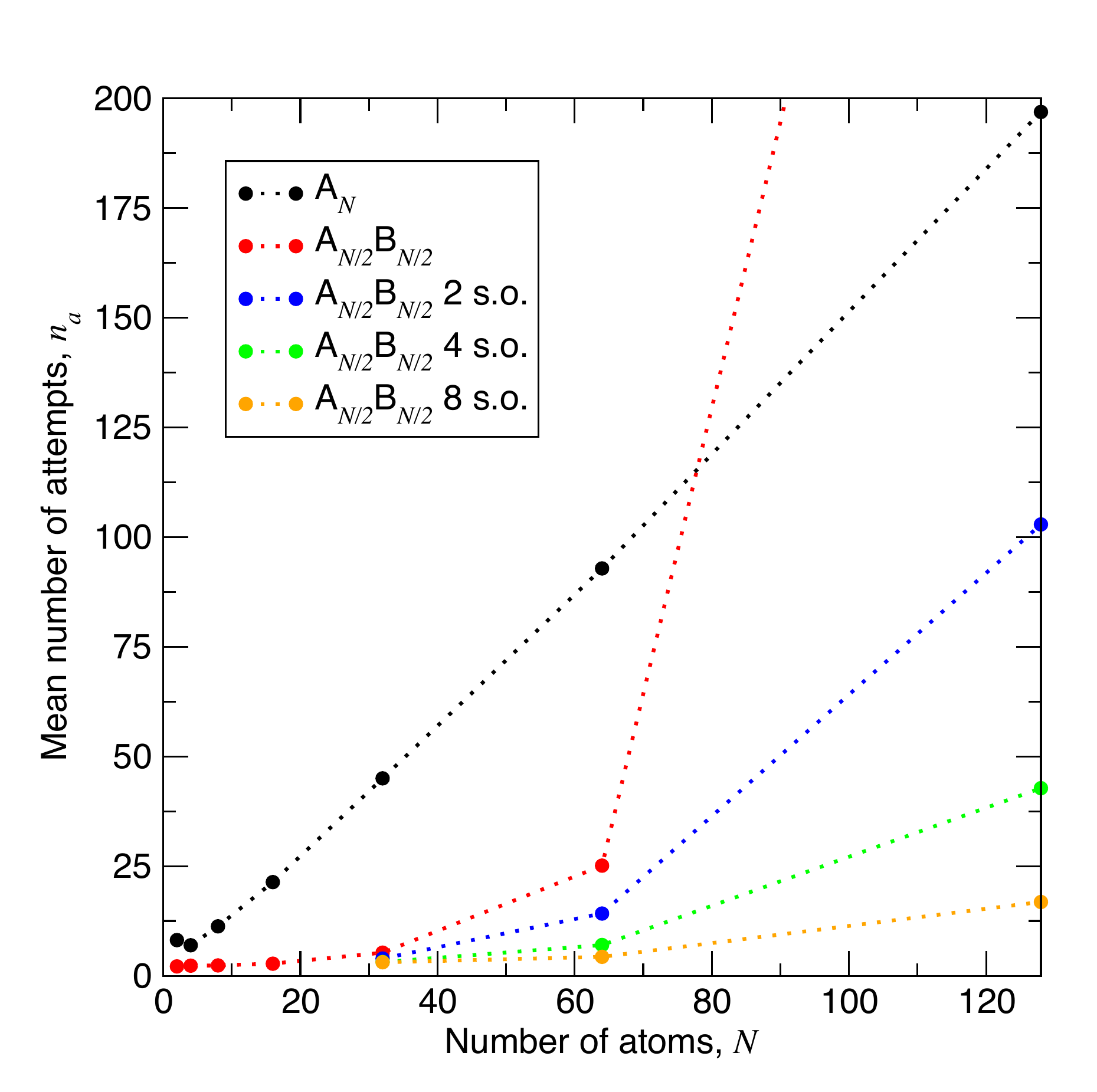}
\caption[]{Data for the mean number of attempts (number of relaxed
  structures) required to find the ground state hcp structure of the
  LJ solid (labelled A$_N$) and the ground state NaCl structure of an
  ``ionic'' solid described by the inter-atomic potential of equation
  (\ref{eq:ionic_potential}) (labelled A$_{N/2}$B$_{N/2}$). We used
  random starting structures and unconstrained searching for the LJ
  solid.  The ionic solid was studied with unconstrained searching,
  and with symmetry constraints consisting of 2, 4, or 8 space-group
  symmetry operations (s.o.).  Structures with 2 space group
  operations were generated by (i) choosing a space group randomly
  from those with 2 operations (ii) choosing the positions of $N/2$
  atoms randomly and generating the positions of the other atoms using
  the symmetry operations.  For both the LJ and ionic potentials, the
  initial cell volumes were renormalised to a random value within
  $\pm$50\% of the equilibrium volumes of the global minimum energy
  structures. The 66\% confidence intervals correspond to about
  $\pm$10 for the largest values of $n_a$ and much smaller
  uncertainties for the other data.}
\label{fig:lennard_jones_solid}
\end{figure}

This review is concerned with searching for structures using
first-principles electronic structure calculations, but the much lower
cost of computing with simple inter-atomic potentials allows more
detailed investigations of the PES and searching algorithms.  Figure
\ref{fig:lennard_jones_solid} shows data for the variation of the mean
number of attempts required to find the global minimum-energy
structure, $n_a$, with the number of atoms per unit cell, $N$.  The
mean numbers of attempts were obtained by relaxing many structures.
The probability $P(n)$ of first obtaining the global minimum after $n$
attempts follows a geometric distribution which, if the probability of
obtaining the global minimum in one attempt is small, tends to an
exponential distribution,
\begin{equation}
  P(n) \simeq \frac{1}{n_a} \exp(-n/n_a).
\label{eq:prob_of_finding_structure}
\end{equation}
The mean number of attempts required to find the global minimum is
therefore
\begin{equation}
  \int_0^{\infty} n \, P(n)\, dn = n_a.
\label{eq:mean}
\end{equation}
The variance of the mean is 
\begin{equation}
  \int_0^{\infty} n^2 \, P(n) \,dn - \left(\int_0^{\infty} n \, P(n) \,dn \right)^2 = n_a^2,
\label{eq:variance}
\end{equation}
which is large when it takes many attempts to find the global minimum.

We have investigated the PES of systems described by some simple
inter-atomic potentials.  We first studied the LJ potential which has
been widely studied as a model of weakly interacting atoms.  Results
for LJ systems are universal because the potential contains only an
energy and a length scale.  We searched for the hcp ground state of LJ
solids with different number of atoms $N$ in the unit cell using RSS
(AIRSS without the ``AI'') and relaxing ``random sensible structures''
as described in Section \ref{subsec:generating_random_structures}. Our
results are shown in figure \ref{fig:lennard_jones_solid}, where the
black dots give the mean number of attempts $n_a$ to find the ground
state hcp structure for $N=$ 2, 4, 8, 16, 32, 64, and 128 atoms.  The
value of $n_a$ varies approximately linearly with $N$ up to $N=$ 128.
As well as the hcp ground state, we found the fcc structure and other
stackings of close-packed layers, and structures containing vacancies
and line and planar defects.  Note that it is slightly easier to find
the ground state structure with 4 atoms than with 2.  We also
performed unconstrained searches using the sequence $N=$ 6, 12, 24,
48, and 96 atoms, which gave similar results but with a slightly
smaller slope.

The fraction of times the ground state structure is found by relaxing
randomly chosen structures is equal to the fraction of the total
volume of structure space occupied by the ground state structure.  A
periodic solid with a unit cell volume of about $N a^3$ has a
structure-space volume of roughly $(Na^3)^{N}$.  The almost linear
variation of the LJ A$_{\rm n}$ curve in figure
\ref{fig:lennard_jones_solid} shows that the volume of structure space
occupied by the hcp structure must increase very rapidly with $N$ up
to $N=128$ atoms.  Increasing the number of atoms in the unit cell
allows more freedom to relax into the ground state.  Beyond $N=128$
the mean number of attempts required to find the ground state starts
to increase more rapidly, making it costly to obtain accurate
statistics, although we were able to find the hcp structure with 256
and 512 atoms.  

We speculate that, for $N > 128$, widely separated regions of the unit
cell tend to become independent and the mean number of attempts
required to find the ground state increases rapidly with cell size,
possibly exponentially.  This implies that the fraction of the volume
of structure space occupied by the hcp ground state falls very rapidly
for $N > 128$.  Oganov and Glass \cite{Oganov_glass_2008} tested their
evolutionary algorithm (EA) on LJ solids, but they did not find the
hcp ground state structure with 256 or 512 atoms.

We have studied the performance of RSS using other potentials which
have different inter-atomic bonding and more than one atomic species.
We considered an AB compound with inter-atomic potential
\cite{Pickard_2010_aluminium}
\begin{equation}
\label{eq:ionic_potential}
V_{ij}(r) = \left(\frac{\sigma_{ij}}{r_{ij}}
  \right)^{12} - \beta_{ij} \left(\frac{\sigma_{ij}}{r_{ij}}
  \right)^{6} ,
\end{equation}
where $\sigma_{\rm AA}$ = 2, $\sigma_{\rm BB}$ = 1.75, 
$\sigma_{\rm AB}$ = (2+1.75)/2, $\beta_{\rm AA}$ = -1, 
$\beta_{\rm BB}$ = -1, $\beta_{\rm AB}$ = 1.

This potential models an ionic compound whose ground state has the
NaCl structure.  

The LJ potential and the ``ionic'' LJ potential of equation
(\ref{eq:ionic_potential}) were truncated at particle separations of
2.5 $\sigma_{ij}$.  Figure \ref{fig:lennard_jones_solid} shows data
for unconstrained searches and for constrained searches where we
impose 2, 4, or 8 symmetry operations.  The ground state NaCl
structure is found with very few attempts at small $N$, whether or not
symmetry is imposed but, for the unconstrained search, it ``takes
off'' very rapidly at larger $N$.  Imposing symmetry constraints
dramatically reduces $n_a$ at large $N$, so that the global minimum
can on average be found in many fewer attempts.  The values of $n_a$
obtained for the ionic system with $N < 64$ are much smaller than for
the LJ system.  We speculate that this is because structures with
adjacent atoms of the same type are likely to be unstable, so that the
number of stable structures is smaller than for the LJ system.  The
value of $n_a$ increases rapidly for $N > 64$.  We believe this arises
from the strong repulsion between like atoms, which reduces the
freedom of the structures to relax, implying that the volume of
structure-space occupied by the ground state is increasing very much
less rapidly than the size of the structure space itself.

We saw that it was rather easy to find the global minimum of the LJ
solid, but it turns out that clusters are much more challenging
because they exhibit geometrical frustration due to the presence of
surface and bulk material.  We performed searches for the LJ26, 38,
55, 75, 98, 100, and 150 clusters.  LJ26 and 55 are easy systems, LJ38
and 75 have double funnels and are more difficult systems, LJ98 is
also a difficult system, and while LJ100 and LJ150 do not have
intrinsically difficult PES, they are quite large.  We were able to
find the global minima of LJ26, 38, and 55 fairly easily with RSS and,
with some effort, LJ100, but we did not find the global minima of
LJ75, 98, or 150.

RSS is the simplest possible searching algorithm as it contains no
variable parameters.  Parameters can be added to the algorithm and
their values optimised for a particular set of systems.  The modified
algorithm may work efficiently for these systems and others, but the
cost of optimising the parameters can be large.  Our aim is therefore
to choose a few variable parameters which give substantial efficiency
gains over a wide variety of systems.  A simple two parameter
extension of RSS is to ``shake'' each relaxed structure a number of
times with some mean amplitude, see Section \ref{subsec:shaking}.  If
a lower energy minimum is found then the shake procedure is repeated.
Leary \cite{Leary_2000} has shown that this algorithm works quite well
for LJ clusters.  We chose reasonable values of the two parameters of
this RASH algorithm (see Section \ref{subsec:shaking}) for LJ clusters
and were able to find the global minimum of each cluster listed above,
although lengthy runs were required for LJ75 and 98 and, to a lesser
extent, LJ150.  Adding a third parameter which allows the occasional
acceptance of moves to higher energy minima leads to a basin hopping
algorithm \cite{Li_1987,Wales_1999,Wales_book_2003}.  One can improve
the efficiency for LJ systems by adding more parameters such as
sophisticated types of atomic displacement, but optimising the
parameters will become more costly and the efficiency for other types
of system may well decline.  In our searches we prefer to make
extensive use of symmetry constraints and prepare initial structures
using chemical units etc., as described in Section
\ref{sec:biasing_searches}.  These powerful ideas are widely
applicable and using them does not involve optimising parameters.

\section{Survey of AIRSS calculations to
  date \label{sec:airss_calculations}}

\subsection{{\bf Silane:}} 

\begin{figure}[ht!]
\centering
\includegraphics[width=0.45\textwidth]{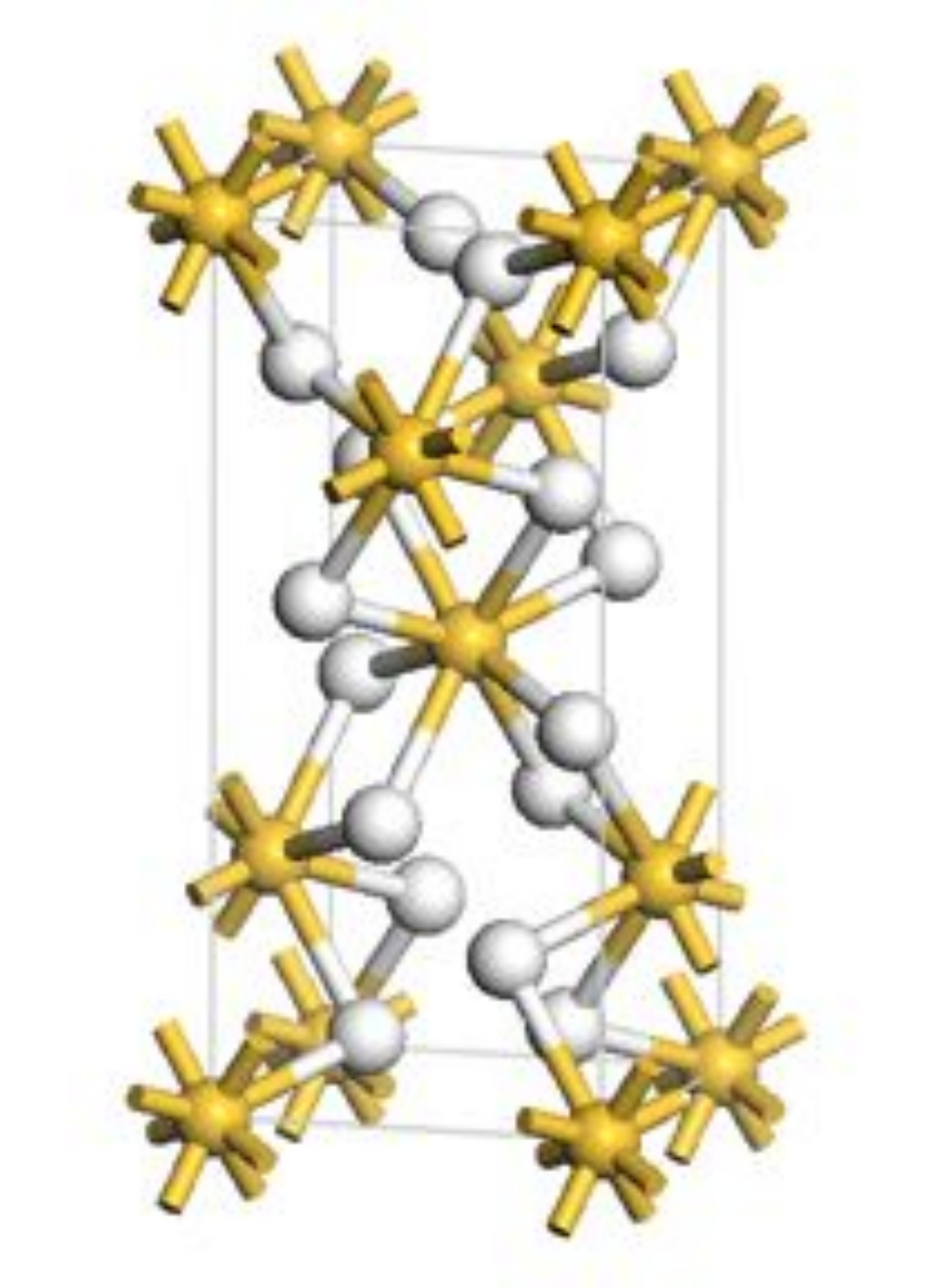}\includegraphics[width=0.45\textwidth]{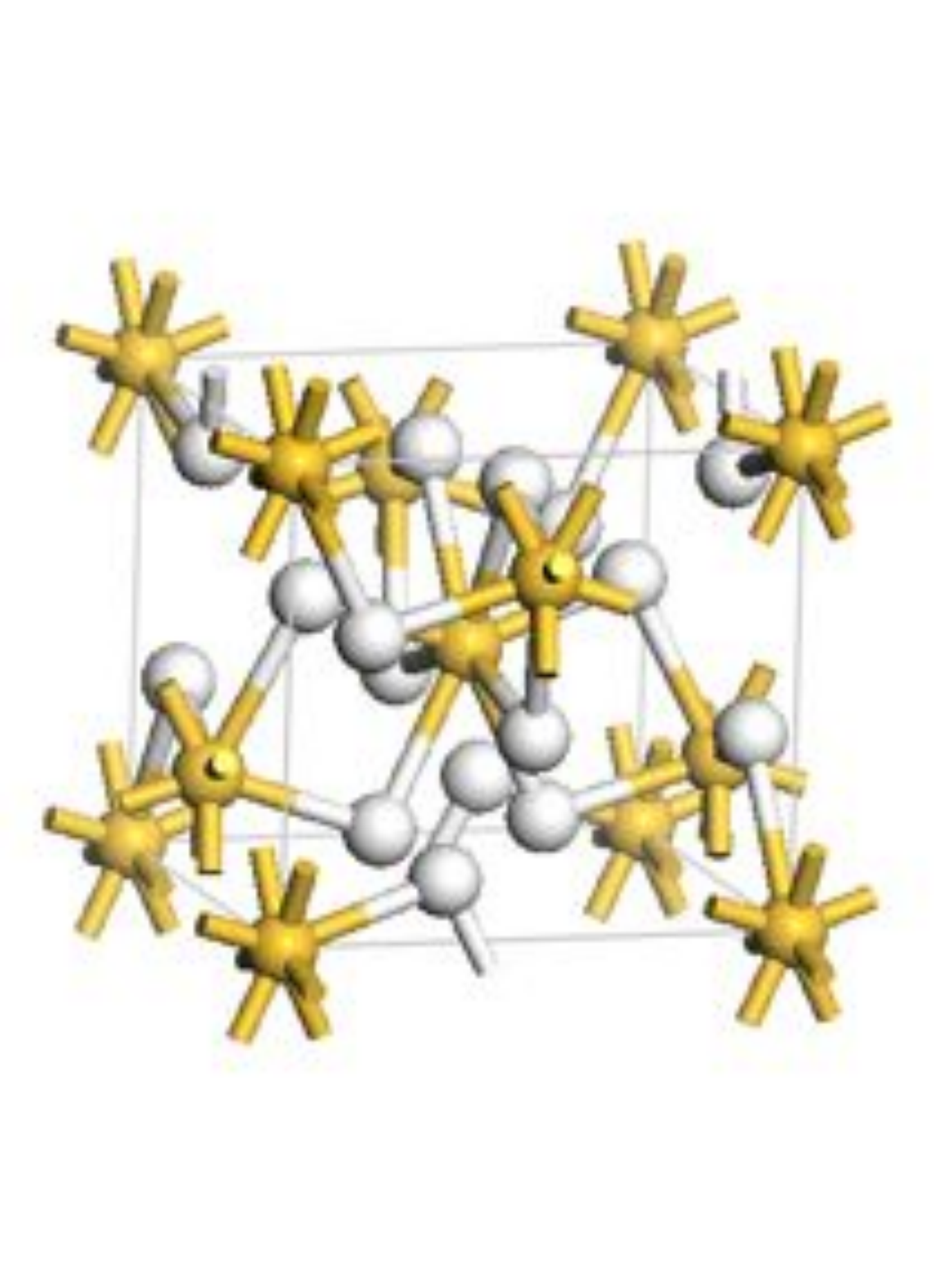}
\caption[]{The $I4_1/a$ structure of silane (left) and the slightly
  less stable $I\bar{4}2d$ structure (right).  Silicon atoms are shown
  in gold and hydrogen atoms are in white.  All of the bonds in
  $I4_1/a$ and $I\bar{4}2d$ are of the Si--H--Si type. Both phases
  were subsequently found experimentally.}
\label{fig:silane}    
\end{figure}

In our first AIRSS paper we studied high pressure phases of silane
(SiH$_4$) \cite{PickardN06}.  This group IVB hydride is a metastable
compound under ambient conditions, but above about 50~GPa it becomes
stable against decomposition into its elements.  Our work was
motivated by a theoretical study \cite{Feng_2006} which used chemical
intuition to predict interesting high pressure non-molecular phases of
silane.  We found more-stable phases, most notably an insulating phase
of $I4_1/a$ symmetry, shown in figure \ref{fig:silane}, which was the
most-stable structure from about 50 GPa to over 200 GPa.  Each Si atom
is bonded to eight H atoms which form bridges between neighbouring Si
atoms.  Each of the Si and H sites are equivalent in this
high-symmetry structure.  All of the bonds are electron-deficient
three-centre-two-electron ``banana'' bonds, similar to those linking
the boron atoms in diborane (B$_2$H$_6$).  Interestingly, Feng
\textit{et al} \cite{Feng_2006} predicted structures with some Si-H-Si
banana bonds, and their chemical intuition was essentially correct,
but our structure is totally bananas.  The $I4_1/a$ phase has
subsequently been observed in x-ray diffraction studies
\cite{Eremets_2008} and its insulating behaviour was verified.  We
also found a slightly-less-stable phase of $I\bar{4}2d$ symmetry only
0.1~eV per SiH$_4$ unit above $I4_1/a$ at 100~GPa.  The $I\bar{4}2d$
phase of silane has also been identified in experiments by Degtyareva
\textit{et al} \cite{Degtyareva_2009}.  An impressive debut for AIRSS!

\subsection{{\bf Aluminium hydride:}}

\begin{figure}[ht!]
\centering
\includegraphics[width=0.5\textwidth]{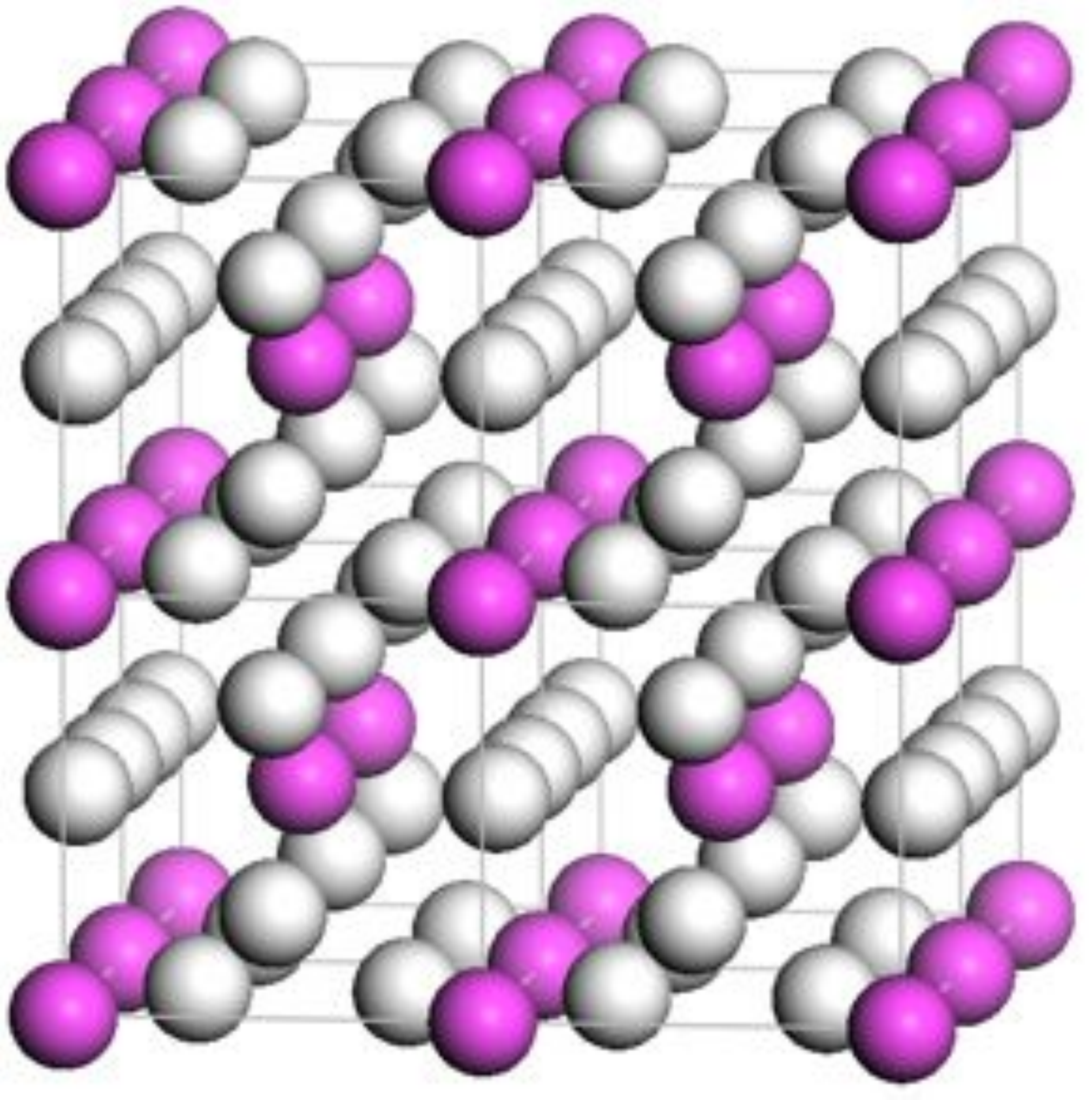}
\caption[]{The $Pm\bar{3}n$ phase of aluminium hydride.  The Al
  cations are shown in purple and the H anions are in white.  The
  linear chains of H atoms can clearly be seen.  This structure is
  also adopted by niobium stannide (Nb$_3$Sn) which is a
  superconductor used in high magnetic field applications.}
\label{fig:Pm-3n_AlH3}    
\end{figure}

The silane studies were
motivated by the quest for metallic hydrogen.  Although metallic
hydrogen has been formed fleetingly in shock wave experiments and must
exist within planets such as Jupiter, it has not been produced in
static compression experiments, where it could be studied in detail.
Hydrides have been thought of as containing ``chemically
pre-compressed'' hydrogen which might become metallic at pressures
achievable in diamond anvil cells and might exhibit phonon-mediated
high-temperature superconductivity \cite{Ashcroft04}.  The group IVB
hydrides contain 80\% hydrogen atoms, but the group IIIB hydrides
contain nearly as much (75\%).  We studied aluminium hydride (AlH$_3$)
and predicted the stability of a metallic $Pm\bar{3}n$ phase at
pressures readily achievable in diamond anvil cells
\cite{PickardN07b}.  The structure of the $Pm\bar{3}n$ phase is
illustrated in figure \ref{fig:Pm-3n_AlH3}.  Hydrogen atoms are
considerably more electronegative than aluminium ones, so the electron
density on the hydrogen atoms is large, which suggests that the
high-frequency hydrogen-derived phonon modes could provide substantial
electron-phonon coupling and promote superconductivity.  However, the
$Pm\bar{3}n$ phase of AlH$_3$ is a semimetal at the transition
pressure with a relatively small electronic density of states at the
Fermi energy, which strongly militates against superconductivity.
$Pm\bar{3}n$ develops a band gap on further compression but, on the
other hand, reducing the pressure increases the density of states at
the Fermi energy which would promote superconductivity.  The
semi-metallic $Pm\bar{3}n$ phase was subsequently observed in
high-pressure x-ray diffraction experiments
\cite{GoncharenkoEHTAYT08}, but it was not found to be a
superconductor.

\subsection{{\bf Hydrogen:}}

\begin{figure}[ht!]
\centering
\includegraphics[width=0.75\textwidth]{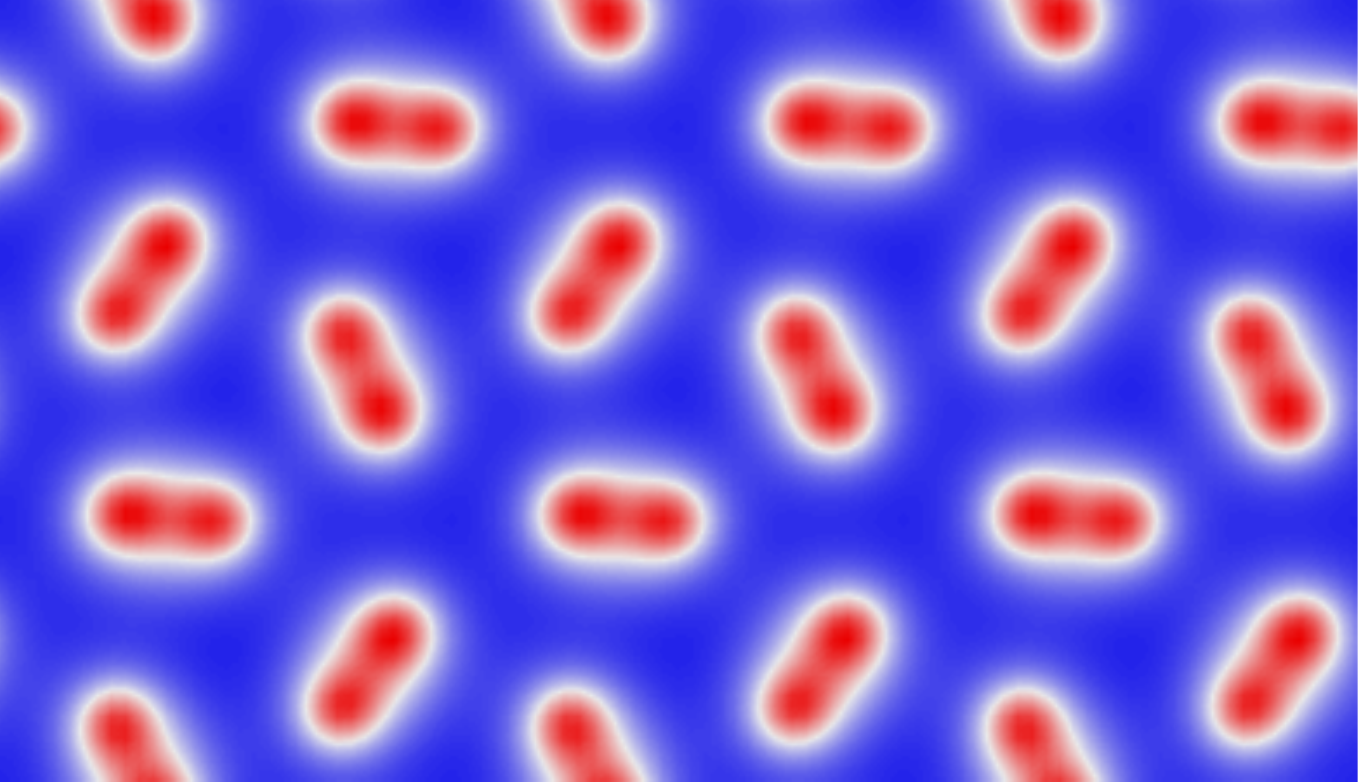}
\caption[]{A slice through the charge density of a layer of the $C2/c$
  molecular hydrogen phase which we predicted to be the most stable in
  the pressure range 105--270 GPa \cite{Pickard_2007_hydrogen}.  Note
  that the two ends of the molecules are inequivalent so they have
  dipole moments and the crystal has infra-red (IR) active vibron
  modes.  The calculations show intense IR vibron activity with strong
  absorption peaks which are close in frequency and would appear as a
  single peak in experiments \cite{Pickard_2007_hydrogen}.  The IR
  activity of the strong IR active vibrons in $C2/c$ increases with
  pressure, as is observed in phase III \cite{HemleyMGM97}.  The
  variation with pressure of the strong IR peak and the Raman active
  vibron frequency of $C2/c$ are in good agreement with experiment
  \cite{GoncharovGHM01}.}
\label{fig:hydrogen}    
\end{figure}

Pure hydrogen has been compressed to over 300 GPa in a diamond anvil
cell \cite{LoubeyreOL02}, but it stubbornly remains insulating.  It is
expected that a non-molecular and presumably metallic phase will
become stable somewhere in the range 400-500 GPa \cite{MaoH94}, and
such pressures will probably be achieved in static experiments in the
near future.  The metallic phase is expected to be a high-temperature
superconductor, perhaps even a room-temperature superconductor.  The
structure of the low-pressure phase I of solid molecular hydrogen is
well established \cite{LoubeyreLHHHMF96}.  Phase II is stable above
110 GPa, and probably consists of molecules arranged on a distorted
close-packed lattice, and a molecular phase III of unknown structure
appears above 150 GPa.  Our AIRSS studies
\cite{Pickard_2007_hydrogen,Pickard_2009_short_review} have shown
there to be several candidate structures for phase II consisting of
packings of molecules on distorted hexagonal-close-packed lattices.
These structures are almost degenerate in enthalpy and quantum motion
of the protons could mean that several significantly different local
molecular configurations contribute to the overall structure of phase
II.  Prior to our work, the DFT phase diagram showed a transition to a
metallic phase below 200 GPa, in strong disagreement with experiment.
We predicted new insulating molecular phases which are stable up to
pressures well above 300 GPa.  In particular, the predicted
vibrational properties of our $C2/c$ molecular phase (which has 24
atoms in the primitive unit cell and is shown in figure
\ref{fig:hydrogen}) agree with the available experimental data for
phase III \cite{Pickard_2007_hydrogen}.

\subsection{{\bf Nitrogen:}} 

\begin{figure}[ht!]
\centering
\includegraphics[width=0.5\textwidth]{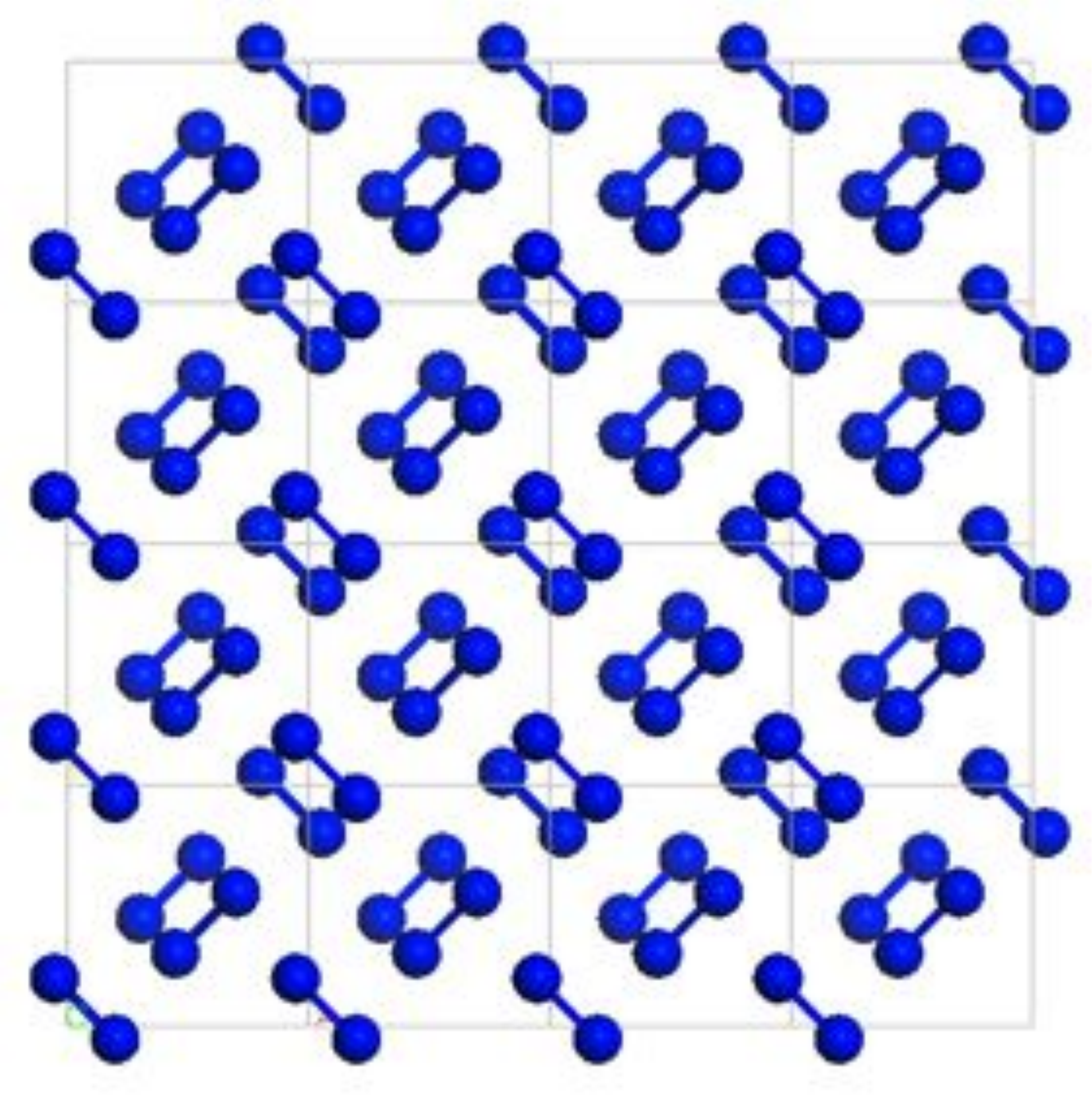}
\caption[]{The $P4_12_12$ molecular phase of nitrogen which we predict
  to be the most stable from $\sim$9.5 GPa up to molecular
  dissociation at about 56 GPa \cite{Pickard_2009_nitrogen}.}
\label{fig:P41212_nitrogen}    
\end{figure}

The phase diagram of nitrogen has been much studied, with a number of
apparently stable and metastable molecular phases having been reported
\cite{BiniUKJ00,GregoryanzGHMSS02,GregoryanzGSSMH07}, although their
structures are mostly unknown.  We found a new class of molecular
structures which we predicted to be more stable than previously
suggested ones over a wide range of pressures
\cite{Pickard_2009_nitrogen}, see figure \ref{fig:P41212_nitrogen}.
The dissociation energy of a nitrogen molecule is more than twice that
of a hydrogen molecule, and yet nitrogen molecules dissociate at far
lower pressures \cite{MartinN86}.  The reason for this is simply that
nitrogen atoms can form up to three covalent bonds so that molecular,
polymeric and dense framework structures are possible, whereas a
hydrogen atom can form only one covalent bond.  The structure of the
high-pressure singly-bonded ``cubic gauche'' phase formed on molecular
dissociation was in fact predicted using DFT calculations
\cite{MailhiotYM92} over a decade before it was observed
experimentally \cite{EremetsGTDB04}, a triumph for chemical intuition.
Computational searches for the phases beyond cubic gauche have also
been performed \cite{Pickard_2009_nitrogen,MaOLXK09}.  Ma \textit{et
  al} \cite{MaOLXK09} used DFT and a genetic algorithm to predict the
phase beyond cubic gauche to be a singly-bonded layered structure of
$Pba2$ symmetry with 16 atoms in the primitive unit cell.  This
structure is slightly more favourable than the very similar
$P\bar{4}2_1m$ structure we found with 8 atoms.  Unfortunately we did
not perform searches with more than 12 atoms, so we could not have
found the $Pba2$ phase.  This serves as a warning to all searchers -
there could always be a better structure in a larger unit cell.

\subsection{{\bf Polymeric Nitrogen:}} 

\begin{figure}[ht!]
\centering
\includegraphics[width=1.0\textwidth]{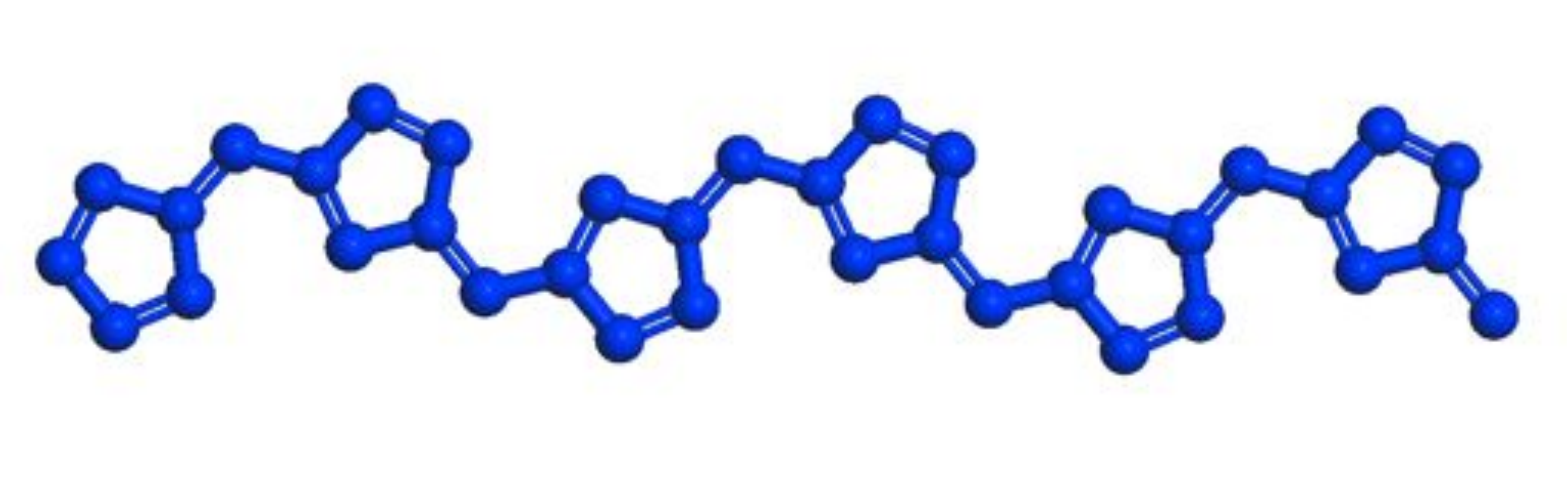}
\caption[]{Nitrogen polymer chain consisting of N$_5$ rings linked by
  a bridging N atom.  Packings of these polymers had the lowest
  energies at zero pressure of all the non-molecular phases of
  nitrogen we found.  }
\label{fig:Cmc21_nitrogen_polymer}    
\end{figure}

The ``polymeric'' nitrogen mentioned above can in fact be recovered to
ambient conditions as a metastable high-energy-density material (HEDM)
\cite{EremetsGTDB04}.  The experimental material is amorphous and is
far too unstable to be a useful HEDM.  An ordered phase could,
however, be significantly more stable.  We therefore set out to find
the most stable non-molecular phase of pure nitrogen at zero pressure.
When we performed searches close to zero pressure we almost always
obtained structures containing N$_2$ molecules, so we searched at 50
GPa instead, where molecular, polymeric and framework structures are
almost degenerate in enthalpy, and then studied the most promising
structures at zero pressure.  Our best structures are polymers
consisting of N$_5$ rings linked by a bridging atom, see figure
\ref{fig:Cmc21_nitrogen_polymer}.  At zero pressure this structure is
about 0.09 eV per N atom lower in energy than the previous best
non-molecular nitrogen crystal structure, which is the $Cmcm$ phase
predicted by Mattson \textit{et al}
\cite{MattsonSCM04,WangTWCLZ_2010}.  The molecular structures are
about 1 eV per atom lower in energy. We calculate our polymer to be
semiconducting with a band gap of about 1.6 eV.  Joining up the ends
of such a nitrogen polymer with four N$_5$N units costs very little
energy and results in a porphin-type structure, which can bind species
such transition metal atoms at its centre.  Alternatively one could
add H atoms to the polymer and porphin-type structures to saturate the
bonding.  The cyclic N$_5^+$ ion, which is the cation of the key
element of our polymer, has already been synthesised, and it does not
seem unreasonable that the polymer or porphin-type structure could be
synthesised.

\subsection{{\bf Water:}}

Our work on structures of H$_2$O \cite{Pickard_2007_water} was
motivated by an experimental study \cite{MaoMMEHCCSH06} in which a new
metastable form of H$_2$O was synthesised.  Mao \textit{et al}
subjected water to an applied pressure of about 20 GPa and 10 keV
x-ray radiation for many hours within a diamond anvil cell, producing
a crystalline phase which does not consist of water molecules.  Mao
\textit{et al} \cite{MaoMMEHCCSH06} concluded that they had
synthesised an alloy of O$_2$ and H$_2$ molecules.  We performed a
AIRSS study at 20 GPa, finding that the structures obtained consisted
almost entirely of weakly bonded H$_2$O, H$_3$O, H$_2$O$_2$,
H$_2$OH$\cdots$OH, H$_2$, and O$_2$ species.  O--H bonds are the most
energetically favourable at 20 GPa, so that the most stable phases
consist of H$_2$O molecules and the highest enthalpy metastable phases
consist of an ``alloy'' of H$_2$ and O$_2$ molecules (rocket fuel!).
We argued \cite{Pickard_2007_water} that the experimental x-ray
diffraction, energy loss, Raman spectroscopy and other data were best
rationalised not by an H$_2$/O$_2$ alloy but by a much more stable
mixture of H$_3$O, O$_2$ and H$_2$ species, no doubt containing
amounts of the other low-enthalpy species.

\subsection{{\bf Ammonia:}}

\begin{figure}[ht!]
\centering
\includegraphics[width=0.5\textwidth]{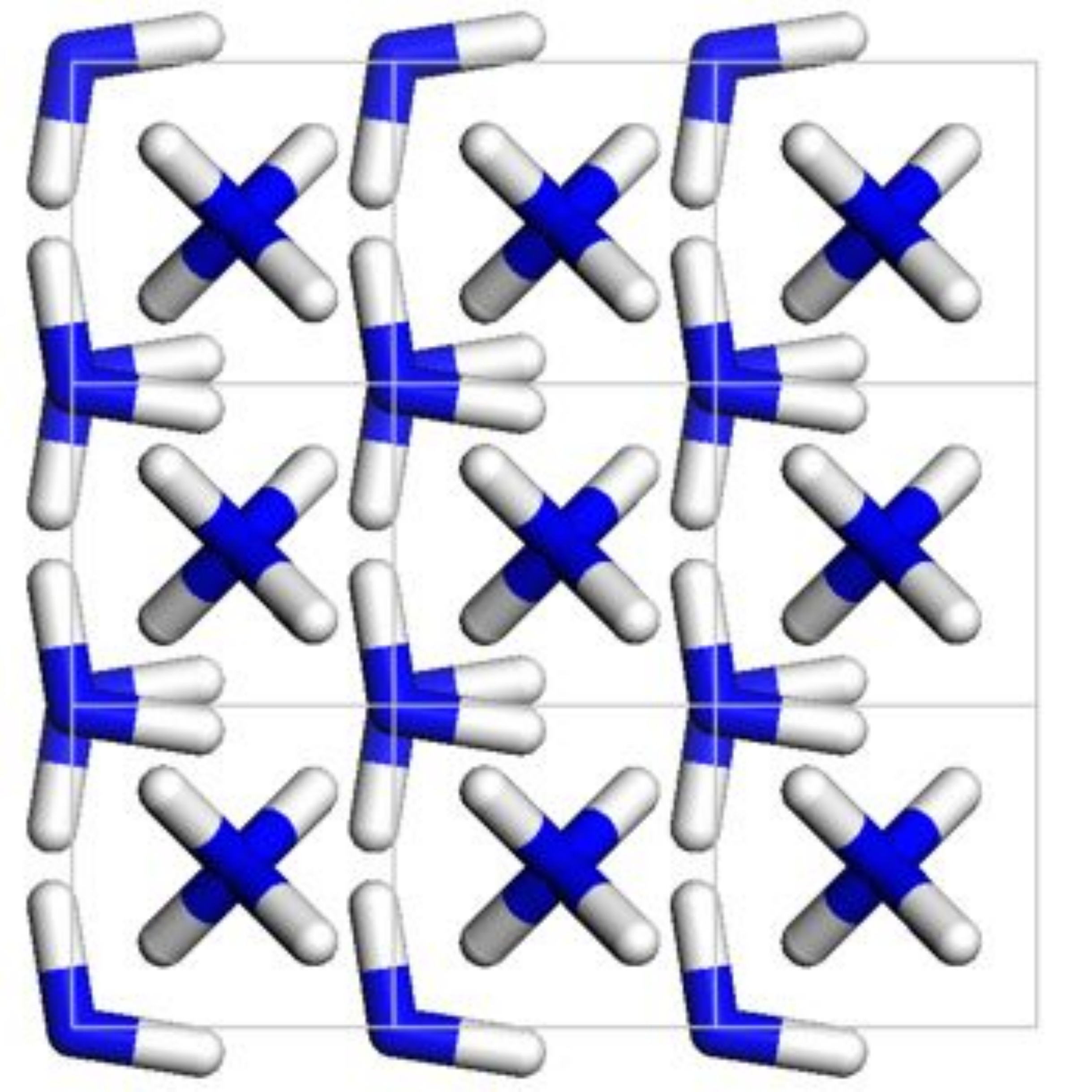}
\caption[]{The ionic $Pma2$ phase of ammonia is predicted to be stable
above 90 GPa and consists of alternate layers of NH$_4^+$ and
NH$_2^-$ ions.  This view shows the three layers of the crystal structure.  
The top layer consists of NH$_2^-$ ions with orientation 
\begin{picture}
(20,10) \put(1,0){\line(0,1){8}}
\put(1,0){\line(1,0){8}} 
\end{picture}
$\!\!\!\!\!\!\!$, the second layer consists of tetrahedrally bonded NH$_4^+$
ions and bottom layer consists of NH$_2^-$ ions with orientation 
\begin{picture}
(20,10) \put(1,8){\line(1,0){8}}
\put(1,8){\line(0,-1){8}} 
\end{picture}
.}
\label{fig:ammonia}    
\end{figure}
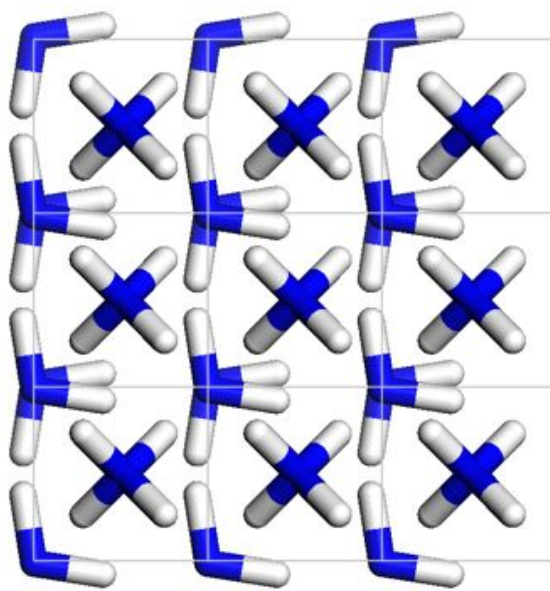

Compressed ammonia (NH$_3$) plays a significant role in planetary
science.  Ammonia forms hydrogen-bonded solids at low pressures, but
we predict that at high pressures it will form ammonium amide ionic
solids \cite{PickardN08}.  These structures, consisting of alternate
layers of ammonium cations (NH$_4^+$) and amide (NH$_2^-$) anions are
expected to be stable over a wide range of pressures readily
obtainable in diamond anvil cells, although experimental verification
of our prediction is still lacking.  The ionic $Pma2$ phase, which is
illustrated in figure \ref{fig:ammonia}, is predicted to be stable
above 90 GPa.  The driving force for the proton transfer reaction is
that the ionic solid is substantially denser than the molecular one.
The proton transfer costs energy under ambient conditions, but at high
pressures the cost is overcome by the lower value of the $pV$ term in
the enthalpy.  A proton transfer between water molecules, forming
OH$^-$ and H$_3$O$^+$ ions, costs more energy than in ammonia and
water molecules pack better than ammonia molecules, so that proton
transfer is not predicted to occur in compressed water.  Proton
transfer is even more favourable in water/ammonia mixtures which are
expected to form OH$^-$ and NH$_4^+$ ions at moderate pressures
\cite{FortesBWVJ01}.

\subsection{{\bf Ammonia monohydrate:} \label{Ammonia monohydrate}}

\begin{figure}[ht!]
\centering
\includegraphics[width=0.75\textwidth]{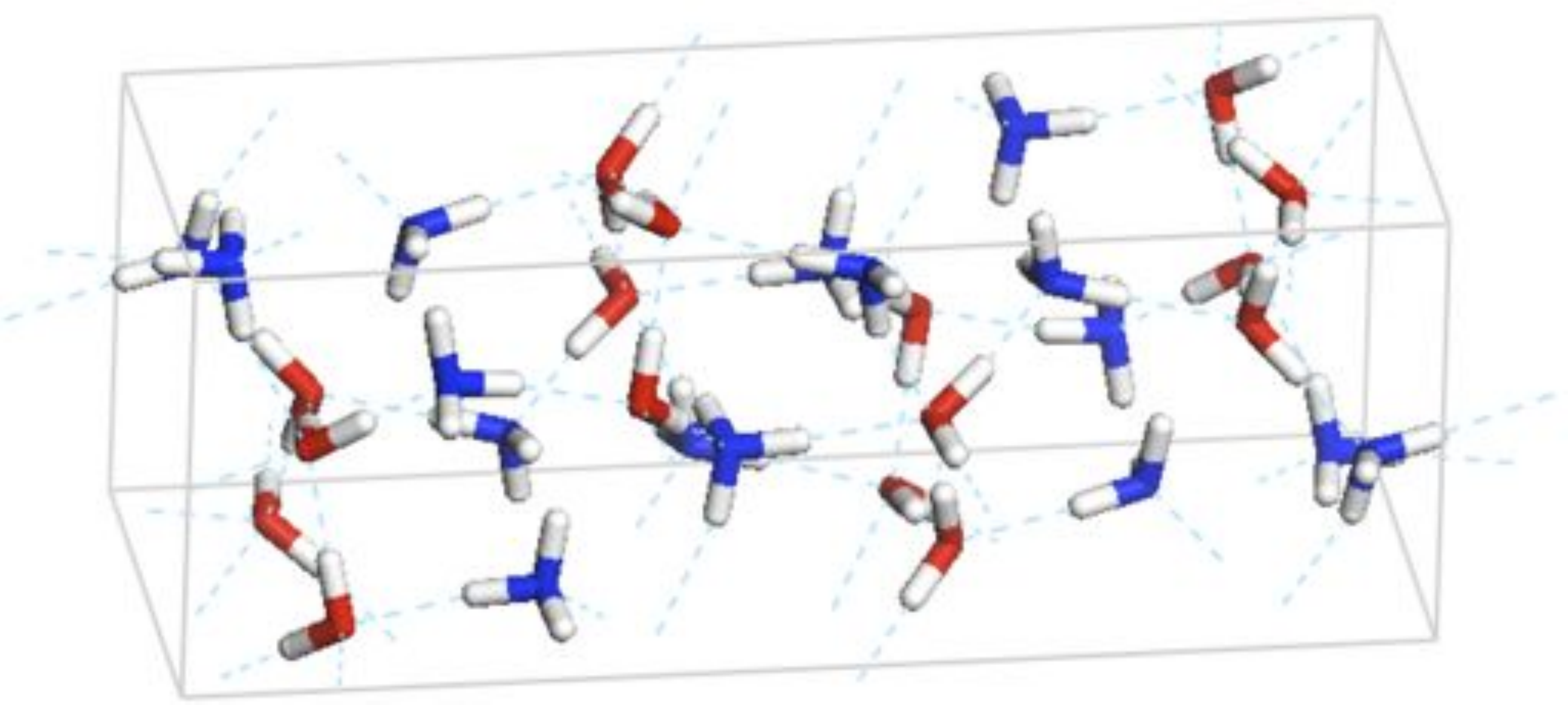}
\caption[]{The structure of phase II of ammonia monohydrate predicted
  using AIRSS.  Oxygen atoms are shown in red, nitrogen atoms in blue,
  and hydrogen in white.  The dashed lines indicate close contacts
  between the molecules.  The structure illustrated above and the
  structure obtained from the neutron diffraction data are very
  similar and both have $Pbca$ symmetry, but they have slightly
  different proton orderings.}
\label{fig:AMH}    
\end{figure}

The properties of compressed ammonia monohydrate
(NH$_3\cdot\mathrm{H}_2$O) are of direct relevance to models of the
formation of Titan, Saturn's largest moon.  Fortes and coworkers
performed neutron diffraction experiments under pressure which yielded
the unit-cell parameters and the candidate space groups ($Pcca$,
$Pnca$ and $Pbca$) of phase II of ammonia monohydrate, which is formed
at pressures of a few tenths of a GPa
\cite{Pickard_2009_structure_ammonia_monohydrate,Pickard_2009_EoS_ammonia_monohydrate}.
The cell parameters indicated that the unit cell contains 16
NH$_3\cdot\mathrm{H}_2$O formula units, giving a total of 112 atoms.
We performed AIRSS calculations using the experimental unit cell with
the further assumption that the crystal consisted of weakly
hydrogen-bonded NH$_3$ and H$_2$O molecules.  Each of the candidate
space-groups contains eight symmetry operations, so the asymmetric
unit contains two formula units.  The initial structures were
generated by inserting two H$_2\mathrm{NH}\cdots\mathrm{OH}_2$ units
at random, generating the rest of the structure using the symmetry
operations and rejecting initial configurations in which the molecules
overlapped strongly.  Searches were performed using each of the three
candidate space groups, and the lowest enthalpy structure was obtained
with space group $Pbca$, see figure \ref{fig:AMH}, which allowed a
refinement based on the original data to be performed.  These results
motivated new experiments which yielded diffraction data which, with
additional insights from our predicted structure, were of sufficient
quality to allow a full structural determination.  A structure of
space group $Pbca$ was determined whose hydrogen bonding network is
almost identical to that of the computationally-derived structure
\cite{Pickard_2009_structure_ammonia_monohydrate,Pickard_2009_EoS_ammonia_monohydrate}.
Subsequent DFT calculations have shown that the experimentally
determined structure is about 0.01 eV per seven-atom formula unit
lower in enthalpy than the theoretically predicted one
\cite{GriffithsNP_2010}.  This project shows the power of constrained
searches.  The size of the parameter space was enormously reduced by
using the cell parameters and candidate space groups from experiment
and the H$_2\mathrm{NH}\cdots\mathrm{OH}_2$ unit assumed on chemical
grounds.  One can never be sure when it is safe to stop searching, and
in this case the search was terminated before the correct structure
was found.  It would certainly have been possible to carry out many
more searches in which the correct structure might well have been
found, but the experimental determination made this redundant.  This
project is a nice example of synergy between experimental and
computational structure determination.
 
\subsection{{\bf Graphite intercalation compounds:}}

Superconductivity was observed in some graphite intercalation
compounds (GICs) in the 1960s.  Interest in GICs was rekindled by the
discovery of substantial superconducting transition temperatures in
C$_6$Ca and C$_6$Yb which increase with pressure
\cite{exp_natphys,exp_emery_herold}. The occupation of an inter-layer
state is correlated with the occurrence of superconductivity
\cite{tcm_natphys}.

\begin{figure}[ht!]
\centering
\includegraphics[width=0.75\textwidth]{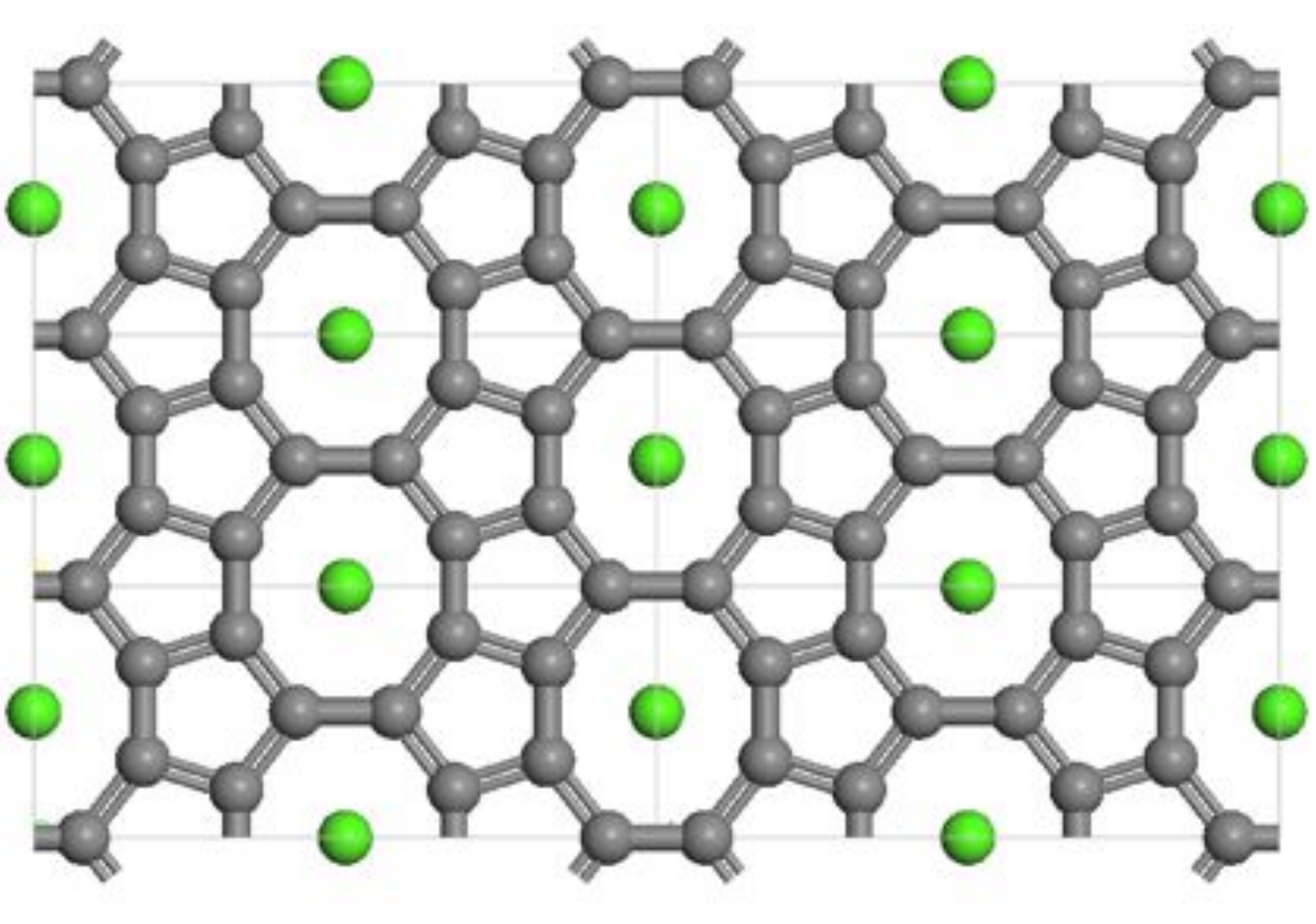}
\caption[]{A CaC$_6$ graphite intercalation compound of $Cmmm$ symmetry.  The
  carbon atoms are shown in grey and the calcium atoms in green.  In the
  $Cmmm$ structure the hexagonal rings of the graphene sheets are replaced by
  five- and eight-membered carbon rings.  This phase is very favourable at
  high pressures because the cost of the Stone-Wales bond rotations is offset
  by a large volume reduction as the metal ions are accommodated within the
  larger rings \cite{Csanyi_2007}.}
\label{fig:Cmmm_gics}    
\end{figure}

Cs\'anyi \textit{et al} \cite{Csanyi_2007} searched for low-enthalpy
structures of C$_6$Ca under pressure.  Energetically competitive
structures were found at low pressures in which the six-membered rings
of the graphene sheets buckle to accommodate Ca atoms within the
troughs.  Stone-Wales bond rotations \cite{Stone} within the graphene
sheets become favourable at higher pressures, leading to structures
with five-, six-, seven- and eight-membered rings, with the Ca atoms
sitting within the larger-diameter rings, see figure
\ref{fig:Cmmm_gics}.  The occurrence of large rings accommodating the
intercalate atoms might be a general features of highly-compressed
GICs, and suggests a route to synthesising novel layered carbon
structures.

\subsection{{\bf Hypothetical group IVB clathrate:}}

\begin{figure}[ht!]
\centering
\includegraphics[width=0.75\textwidth]{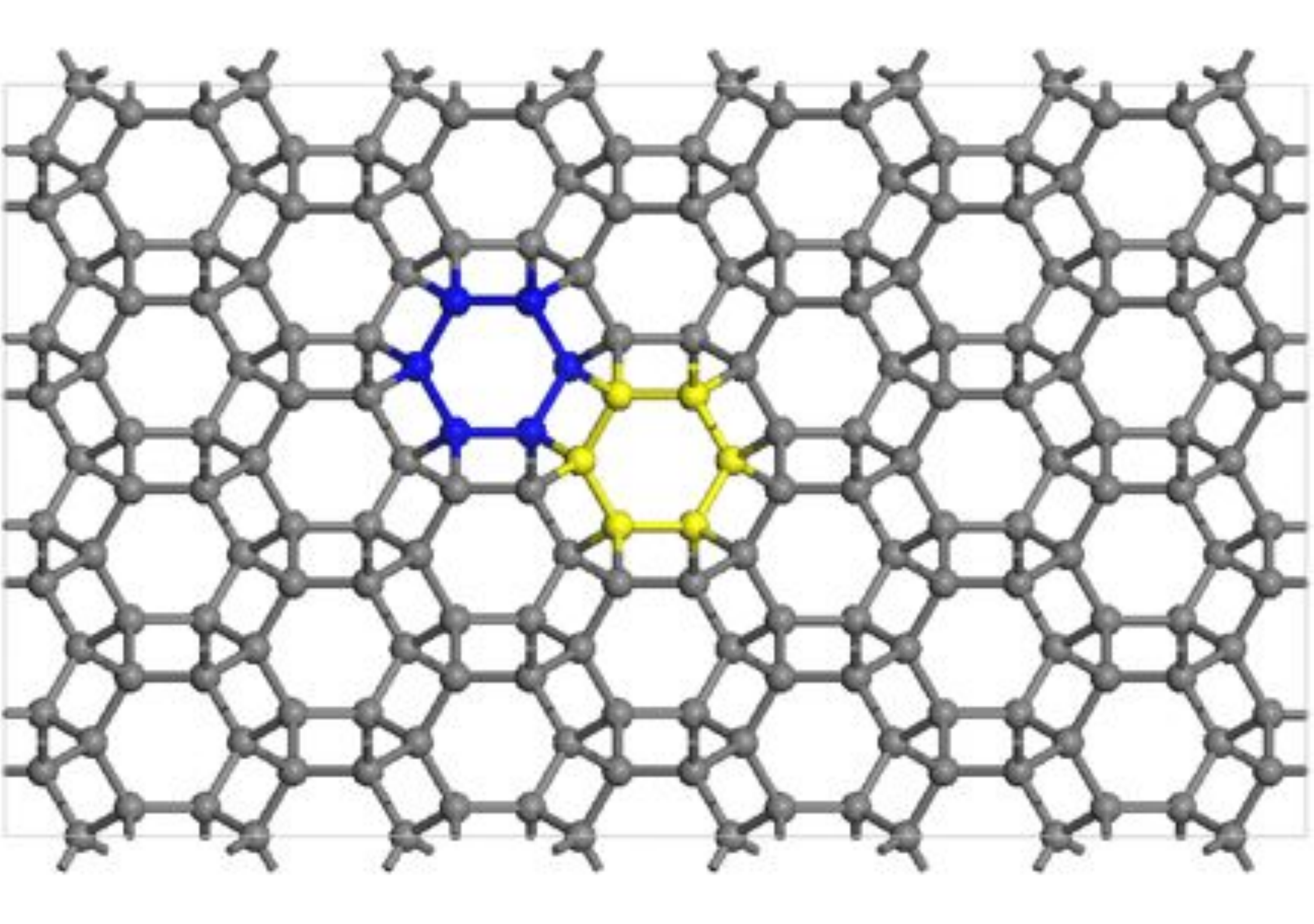}
\caption[]{View of the ``chiral framework structure'' (CFS) along the
  axis of the helices \cite{Pickard_2010_kitchen_towel}.  The CFS is a
  low-energy hypothetical structure of group IVB elements which is
  only a little higher in energy than the diamond structure.  The CFS
  has six atoms per primitive unit cell which are all equivalent by
  symmetry.  The atoms are arranged in five-membered rings and are
  four-fold coordinated.  The CFS has three bond angles slightly
  smaller than the perfect tetrahedral angle of 109.5$^{\circ}$ and
  one bond angle of about 125$^{\circ}$.  The structure consists of a
  hexagonal packing of helices which are crosslinked to satisfy
  four-fold coordination.  The helices all twist either to the left or
  right, so that the crystal is chiral and cannot be superimposed on
  its mirror image.}
\label{fig:kitchen towel}    
\end{figure}

AIRSS produces many structures and the metastable ones are often
interesting.  Looking at the results of a search on carbon we noticed
a low-density high-symmetry $sp^3$-bonded structure which was
unfamiliar to us \cite{Pickard_2010_kitchen_towel}.  This structure
(figure \ref{fig:kitchen towel}) has a six-atom primitive unit cell
with all atoms equivalent, and it is chiral, so that it cannot be
superimposed on its mirror image.  We have named this the ``chiral
framework structure'' (CFS).  It is only 112 meV per atom higher in
energy than carbon diamond, while in silicon it is 53 meV per atom
higher in energy than the diamond structure
\cite{Pickard_2010_kitchen_towel}.  Further investigation revealed it
to be the elemental analogue of a zeolite-type structure and it is
also related to clathrate structures.  Recently we have been made
aware that DFT calculations for the CFS in silicon and germanium had
previously been reported by Conesa \cite{Conesa_2002}, who obtained
similar values for the energy differences from the diamond structures.
Clathrate structures of several different types have been synthesised
consisting of silicon, germanium and tin (but not carbon)
\cite{KasperHPC65,San-MiguelT05}. The synthesis can only be performed
by including ``guest'' atoms such as Na, K, Rb, Cs or Ba, which act as
templates for the self-assembly of the nano cages forming the
structures, although in some cases the guest atoms can largely be
removed. The clathrate II structures of silicon and carbon are
calculated to be about 52 meV per silicon and 72 meV per carbon atom
higher in energy than the corresponding diamond structures.
Considering that the silicon clathrate II structure has been
synthesised \cite{BeekmanN08}, might it be possible to synthesise the
silicon CFS?  A suitable template would have to be found, but it is an
intriguing possibility.

\subsection{{\bf Tellurium dioxide:}}

Metal dioxides with large cation radii often form cotunnite phases
under high pressures, and presumably these transform to post-cotunnite
structures at higher pressures.  Tellurium dioxide (TeO$_2$) is
apparently the only dioxide in which a post-cotunnite phase has been
observed \cite{SatoFYM2005}, and it is therefore a candidate for the
post-cotunnite structure of other metal dioxides.  Unfortunately the
quality of the x-ray diffraction data obtained by Sato \textit{et al}
for post-cotunnite TeO$_2$ was insufficient to allow a structural
determination, although it was possible to eliminate the known
post-cotunnite structures of dihalides \cite{SatoFYM2005}.  Our AIRSS
study \cite{Griffiths_2009_TeO2} found a transition to a
post-cotunnite phase of TeO$_2$ at 130 GPa, which is a little higher
than the experimental transition pressure of 80-100 GPa. The
calculated x-ray diffraction data for the predicted phase of $P2_1/m$
symmetry is in reasonable agreement with experiment.  Interestingly we
found that the cotunnite phase shows re-entrant behaviour, becoming
more stable than $P2_1/m$ again above 260 GPa.  We tried our $P2_1/m$
structure in other metal dioxides but it was never the most stable
phase \cite{Griffiths_2009_TeO2}.  Higher quality x-ray diffraction
data are required to test our identification of the $P2_1/m$ structure
as post-cotunnite TeO$_2$.

\subsection{{\bf Lithium-beryllium alloys:}}

AIRSS was adopted by Feng \textit{et al} \cite{FengHAH2008} for
exploring lithium-beryllium (Li-Be) alloys under pressure.  These
elements are immiscible under ambient conditions, but the calculations
show they can react under pressure, with LiBe$_2$ becoming more stable
than the separated elements above about 15 GPa, and Li$_3$Be, LiBe and
LiBe$_4$ having regions of stability at higher pressures.  The
electronic structure of the most stable LiBe compound shows
two-dimensional character, with a characteristic step-like feature at
the bottom of the valence band.  The changes in the electronic
structure which allow the formation of Li-Be alloys under compression
arise from overlap of the Li $1s$ core electrons which leads to charge
transfer towards the Be atoms.

\begin{figure}[ht!]
\centering
\includegraphics[width=0.75\textwidth]{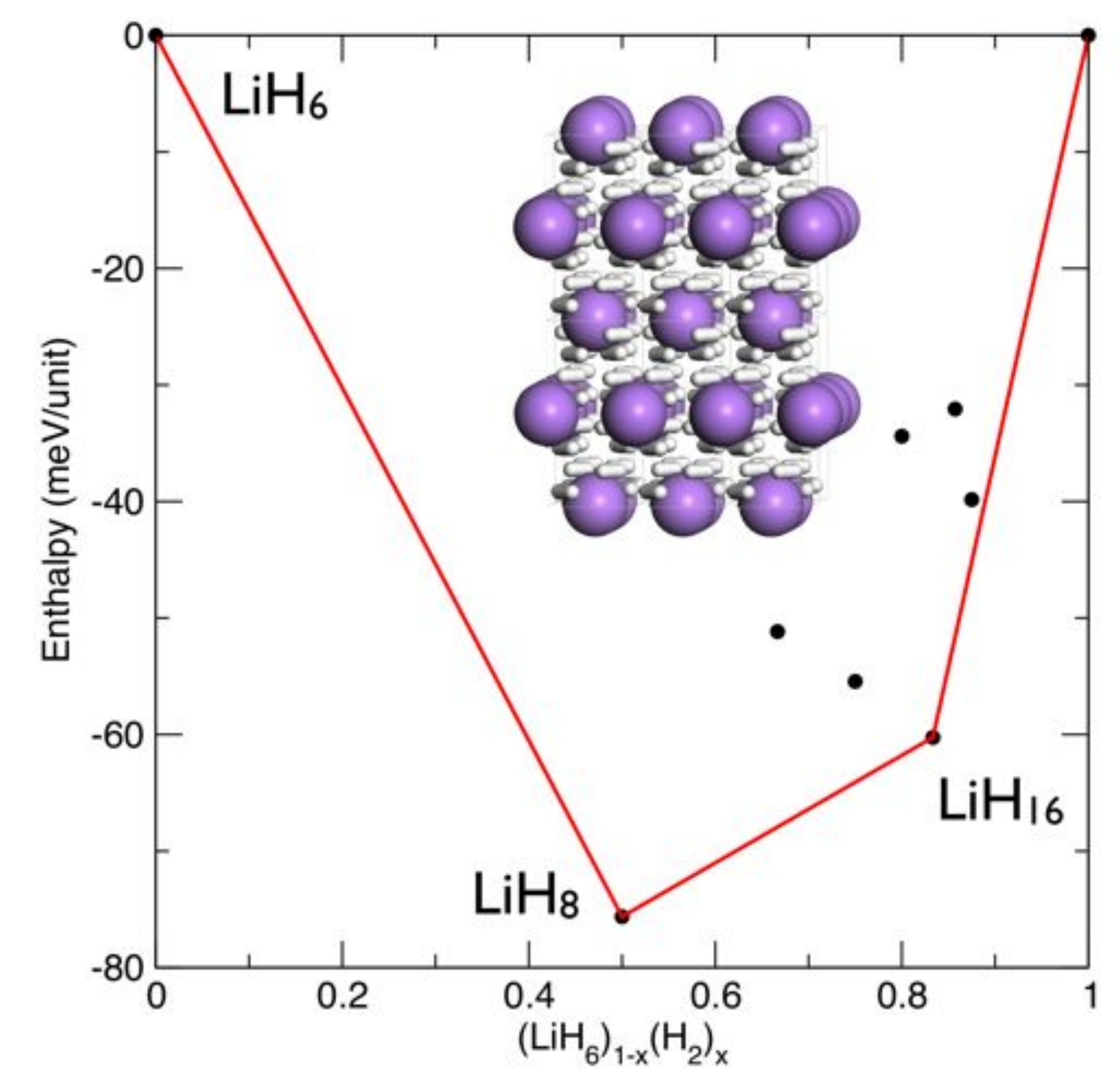}
\caption[]{Zurek \emph{et al} have found that ``a little bit of
  lithium does a lot for hydrogen''\cite{zurek_2009}. In more
  extensive variable stoichiometry searches we find that even less
  lithium will do the trick of ``metallising hydrogen''. We determined
  the low enthalpy structures of LiH$_{2n}$ for $n=3-10$ at 100~GPa
  using AIRSS, displaying the results on a convex hull. LiH$_{16}$
  (shown) is stable against decomposition into LiH$_8$ and H$_2$. It
  is metallic and is based on a body-centred-tetragonal (bct) packing
  of lithium atoms ``coated'' in H$_2$ molecules. The structure of
  LiH$_{16}$ is given in \ref{sec:new_structures}. We have found
  similar structures for Na, K, Rb and Cs at lower pressures.}
\label{fig:Li+H}
\end{figure}

In this work \cite{FengHAH2008} the relative stabilities of the
different stoichiometries was displayed using a ``convex hull''
diagram.  An example of a convex hull diagram constructed using data
obtained from our random searches for the Li-H system is shown in
figure \ref{fig:Li+H}.

\subsection{{\bf Lithium: }}

\begin{figure}[ht!]
\centering
\includegraphics[width=0.6\textwidth]{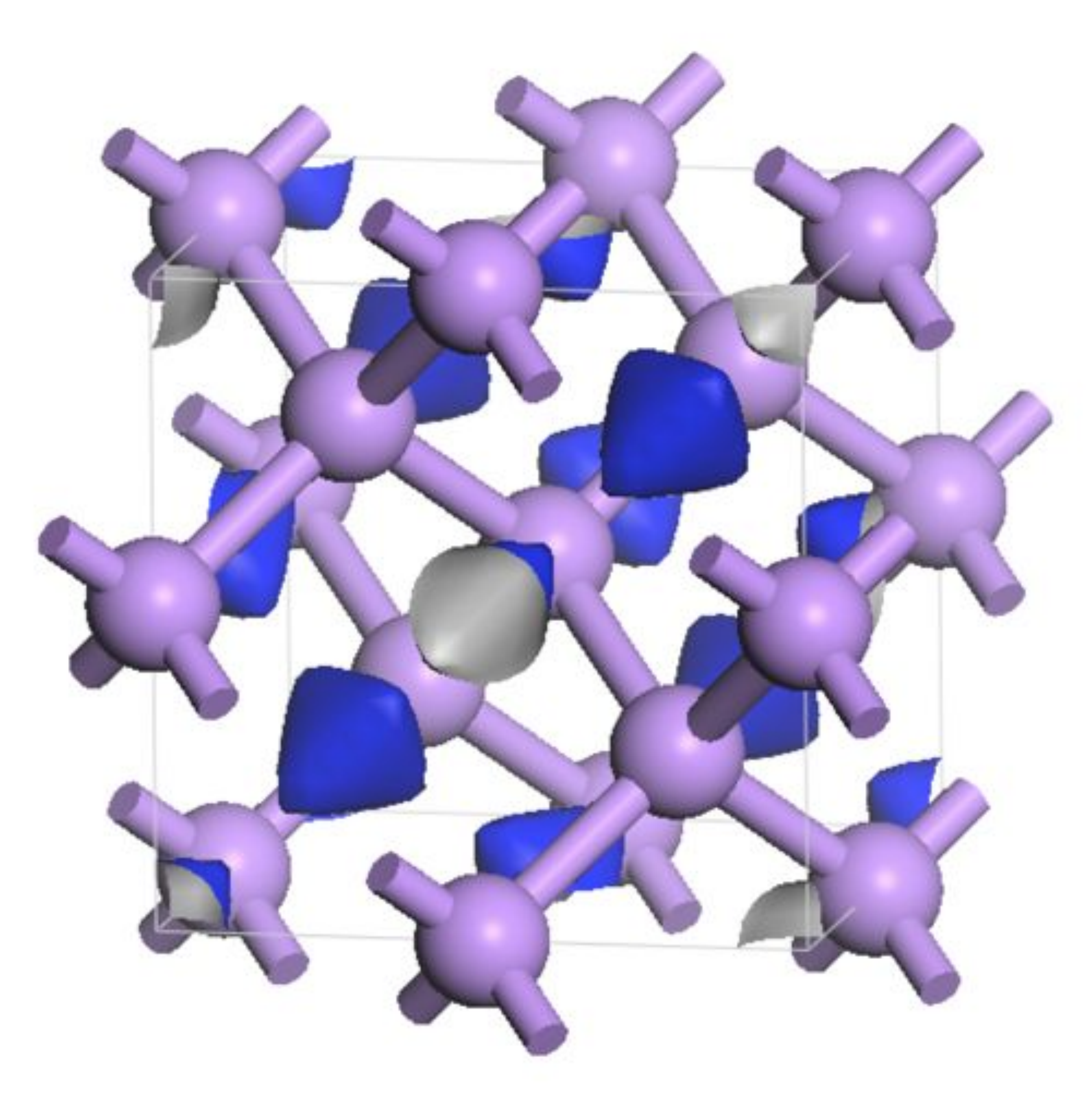}
\caption[]{The diamond-structure electride phase of Li, which is
  predicted to be stable above 383 GPa. The Li atoms are shown as
  purple balls and nearest neighbour contacts are shown as sticks. The
  charge isosurface in blue shows electrons also located on a diamond
  lattice, in the voids between the lithium ions.}
\label{fig:Lithium}    
\end{figure}

One of the surprises in high pressure physics in recent years has been
the discovery that $sp$-bonded elements often adopt complex
non-close-packed structures under sufficient compression.  The ionic
cores take up a larger fraction of the total volume under pressure and
some of the valence charge is pushed away from the atoms and into
interstitial regions forming ``blobs'' which are rather isolated from
one another.  The resulting structure can be thought of as an
``electride'' in which the interstitial electrons are the anions.  The
valence electronic energy bands consequently become narrower than the
free-electron bands \cite{NeatonA99,RousseauA08}.  Lithium (Li) adopts
the fcc structure under ambient conditions, but it transforms to a
three-fold coordinated structure at about 40 GPa \cite{HanflandSCN00}.
We searched for structures of Li at high pressures, finding two new
candidate phases of $Pbca$ and $Aba2$ symmetry which are predicted to
have small regions of stability around 100 GPa
\cite{Pickard_2009_lithium} and are distortions of the $Cmca$-24
structure found in a previous theoretical study \cite{RousseauUKT05}.
All of these structures have substantial dips in their electronic
densities of states (e-DOS) around the Fermi level.  This is
consistent with, but does not fully explain, the significant increase
in electrical resistivity and change in its temperature dependence
near 80 GPa observed by Matsuoka and Shimizu \cite{MatsuokaS209}.  The
occupied valence bandwidths of the $Pbca$, $Aba2$ and $Cmca$-24 phases
are substantially narrower than the corresponding free-electron
values, demonstrating their electride nature.  The low (three-fold)
coordination number of these structures arises from Jahn-Teller-like
distortions which lower the e-DOS around the Fermi level, and we
predicted that the coordination will increase to four-fold above about
450 GPa \cite{Pickard_2009_lithium}, with the diamond structure, see
figure \ref{fig:Lithium}, becoming stable above $\sim$500 GPa.  A
first-principles study was also performed by Yao \textit{et al}
\cite{YaoTK09}, who found similar results using random structure
searching and an evolutionary algorithm.  Overall we are, however,
left with the impression that there are many nearly-degenerate
structures around 100 GPa, and more twists in the story of compressed
Li are likely.

\subsection{{\bf Boron:}}

Under ambient conditions the $\alpha$ and $\beta$ phases of boron are
almost degenerate in energy, but $\alpha$-boron is more dense and is
favoured at higher pressures.  The structure of a high-pressure
$\gamma$-phase of boron was recently solved by combining x-ray
diffraction data with an evolutionary structure-prediction algorithm
using first-principles calculations
\cite{Solozhenko_boron_2008,Oganov_boron_2009}, and from x-ray
diffraction data alone
\cite{Zarechnaya_boron_2008,Zarechnaya_boron_2009}.  The
$Pnnm$-symmetry $\gamma$ phase has 28 atoms in the primitive unit cell
and consists of B$_{12}$ icosahedra and B$_{2}$ dimers.  The $\alpha$
phase was found to be unstable to the $\gamma$ phase above 19 GPa, and
DFT calculations show that the $\gamma$ phase gives way to the
$\alpha$-Ga structure of $Cmca$ symmetry above 89 GPa
\cite{Haussermann_boron_2003}.

\begin{figure}[ht!]
\centering
\includegraphics[width=0.35\textwidth]{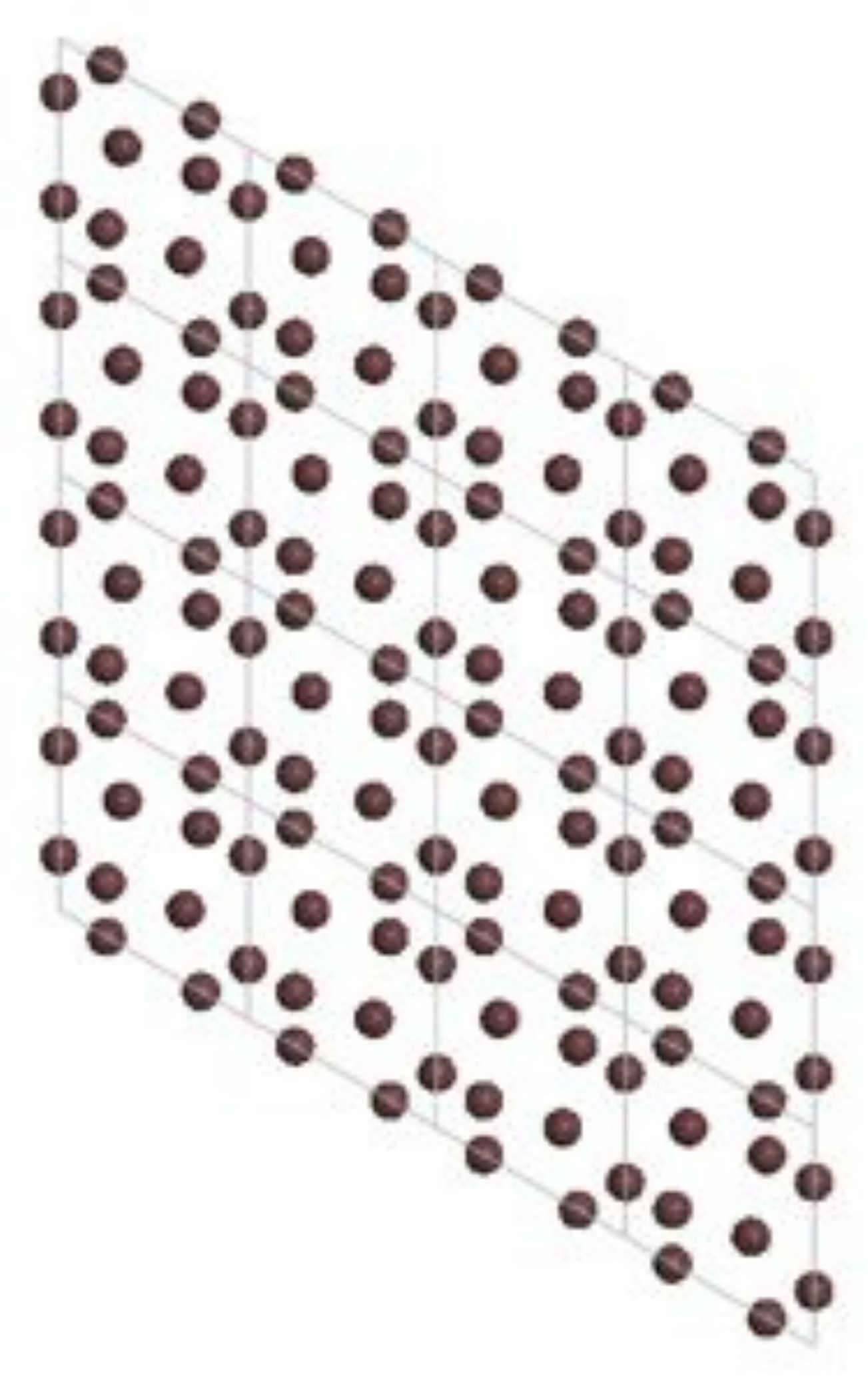}\hspace{1cm}\includegraphics[width=0.35\textwidth]{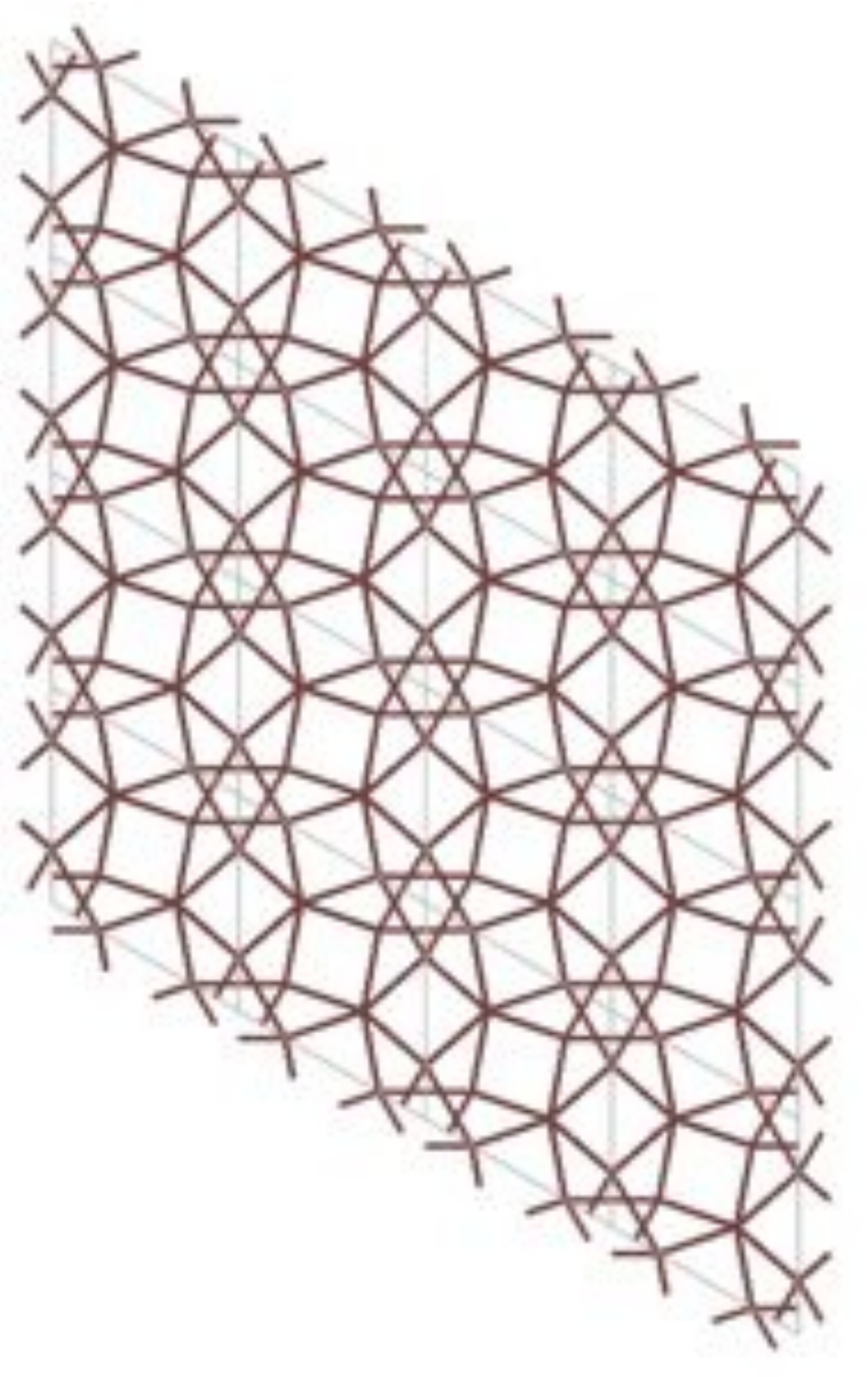}
\caption[]{The $P6_3/mcm$ ``host-guest'' structure of boron, which we
  predict to be the most stable phase above 383 GPa.  The structure is
  that of the Mn atoms in the Mn$_5$Si$_3$Z ternary compound.  The
  picture on the left shows the positions of the boron atoms in
  $P6_3/mcm$ viewed along the guest chains, and the picture on the
  right shows the ``bonds'' or close contacts between atoms.  The host
  atoms form tubes which can be seen in the figure as hexagons while
  the other atoms form guest chains. }
\label{fig:boron_P63mcm}    
\end{figure}

We have searched for the phase beyond $Cmca$ and found a structure of
$P6_3/mcm$ symmetry with 10 atoms per primitive unit cell which we
find to be the most stable phase above 383 GPa.  The volume of
$P6_3/mcm$ is 3.5\% smaller than $Cmca$ at the transition, and
therefore it rapidly becomes more stable at higher pressures.  The
atoms of the metallic $P6_3/mcm$ host-guest structure occupy the Mn
sites of the Mn$_5$Si$_3$Z ternary compound, where Z can be a variety
of atoms \cite{Corbett_1998,Rogera_2006}, as shown in figure
\ref{fig:boron_P63mcm}.  The structure is of electride type with the
valence electrons sitting on the Si and Zn sites.  We found that
varying the number of atoms in the guest chains increased the
enthalpy, so this phase of boron is locked into a commensurate
structure, although it might be incommensurate in another element.  A
phase of symmetry $I4/mcm$ with 10 atoms in the primitive cell is
found to be about 0.1 eV/atom less stable than $P6_3/mcm$. $I4/mcm$ is
also a host-guest electride based on the W$_5$Si$_3$ binary compound,
and it is the commensurate analogue of the host-guest phase found in
aluminium \cite{Pickard_2010_aluminium}.  

\begin{figure}[ht!]
\centering
\includegraphics[width=0.7\textwidth]{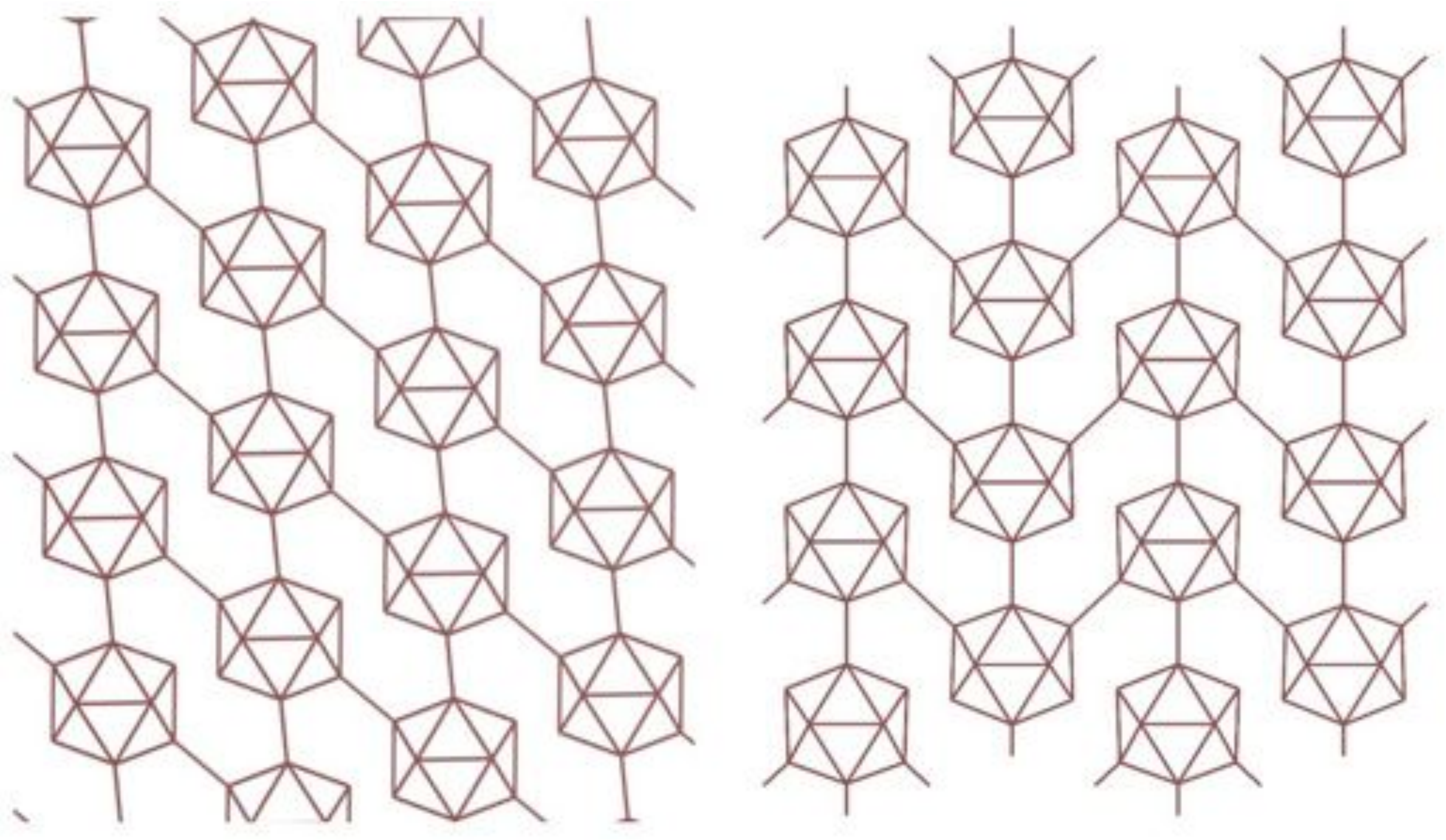}
\caption[]{The $\alpha$ and low-pressure $Cmcm$ structures of boron at
  10 GPa. The low-pressure $Cmcm$ structure is only slightly higher in
  enthalpy than $\alpha$-boron. }
\label{fig:alpha_and_Cmcm_boron}    
\end{figure}

\begin{figure}[ht!]
\centering
\includegraphics[width=0.35\textwidth]{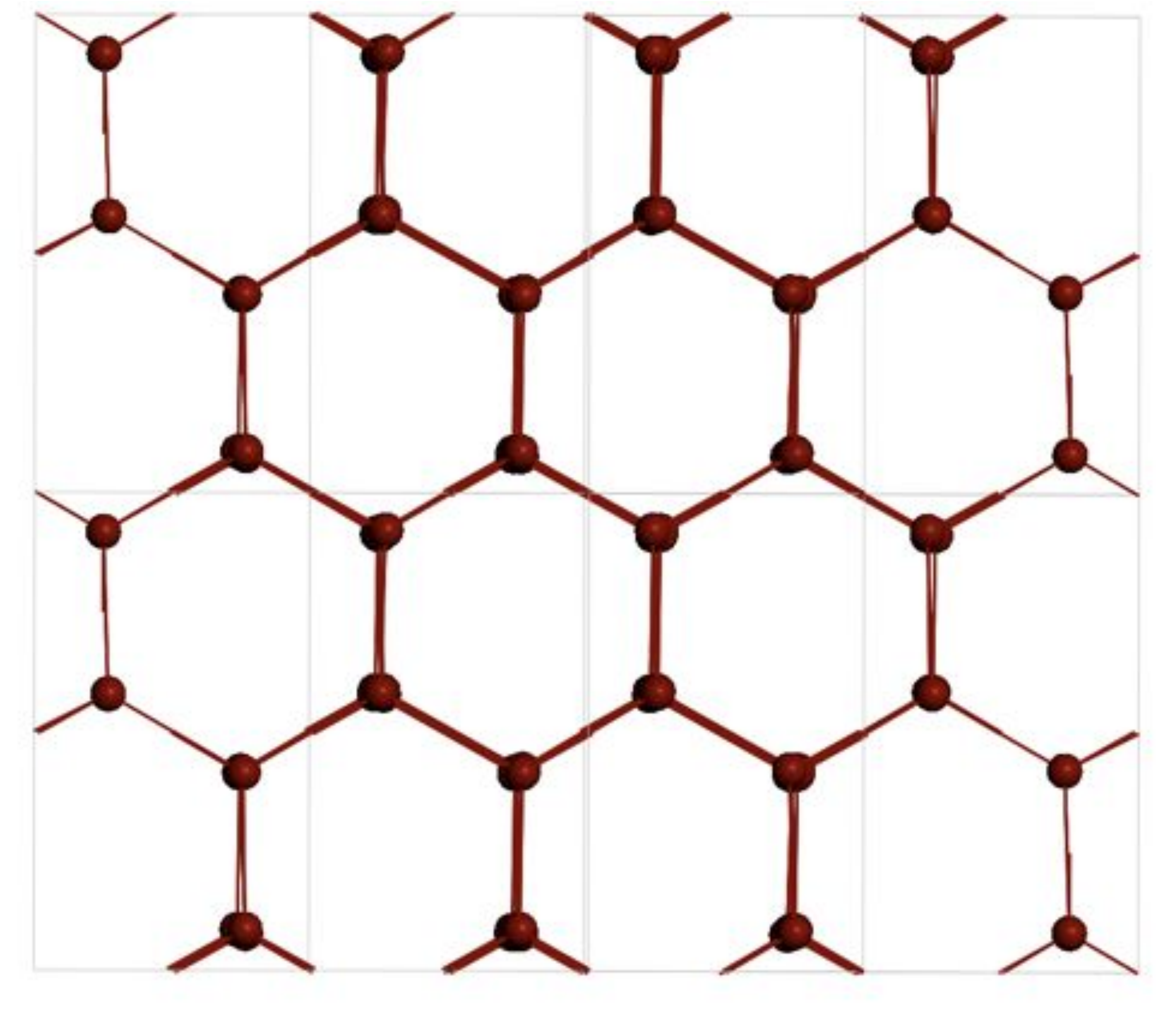}\hspace{0.5cm}\includegraphics[width=0.35\textwidth]{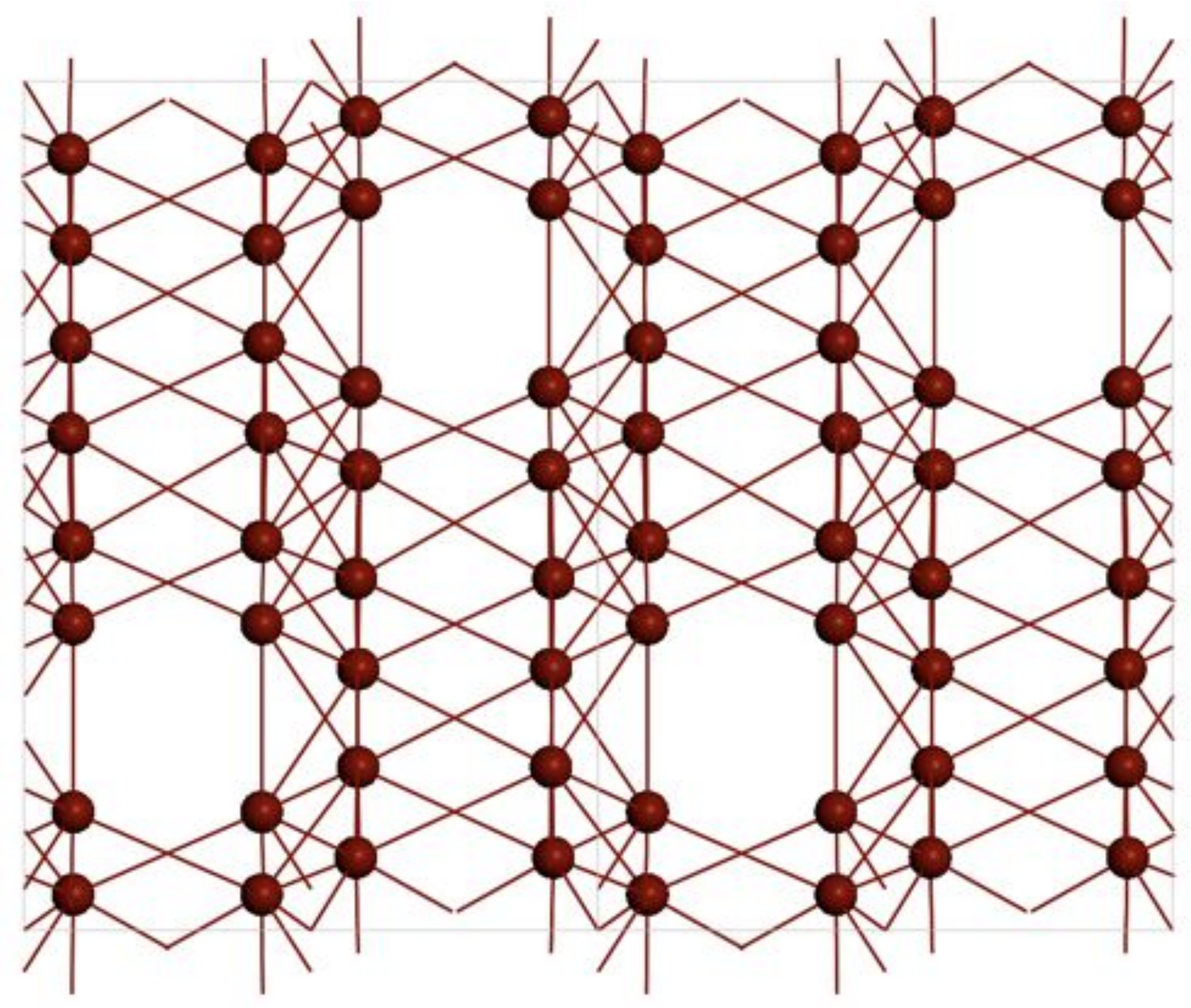}
\caption[]{End and side views of a member of a family of boron
  structures with densities intermediate between $\gamma$-boron and
  the high-pressure $Cmca$ phase.  These phases are very close in
  enthalpy to $\gamma$-boron and high-pressure $Cmca$ around 90 GPa,
  but they are never the most stable.  They are based on various
  supercells of a structure of space group $P6_3/mmc$, c/a$\simeq$0.6
  and atoms on the (2/3,1/3,1/4) Wyckoff position, giving two atoms
  per primitive cell.  So as to satisfy electron counting rules,
  either one atom in nine is missing, or planar defects are present.
  The example illustrated here has 16 atoms per primitive cell and
  space group $C2/c$.}
\label{fig:boron_B2n}    
\end{figure}

It appears that there are many energetically competitive phases in
boron, and we have uncovered a number of metastable phases at lower
pressures.  We found a $Cmcm$ structure illustrated in figure
\ref{fig:alpha_and_Cmcm_boron}, which is a polymorph of
$\alpha$-boron, differing in the connectivity of the icosahedral units
and just 0.01 eV/atom less stable.  We also found a family of
structures more dense than $\gamma$-boron but less dense than $Cmca$,
which commonly appeared in our searches, an example of which with 16
atoms in the primitive cell is shown in figure \ref{fig:boron_B2n}.

\subsection{{\bf Aluminium:} \label{subsec:Aluminium}}

\begin{figure}[ht!]
\centering
\includegraphics[width=0.75\textwidth]{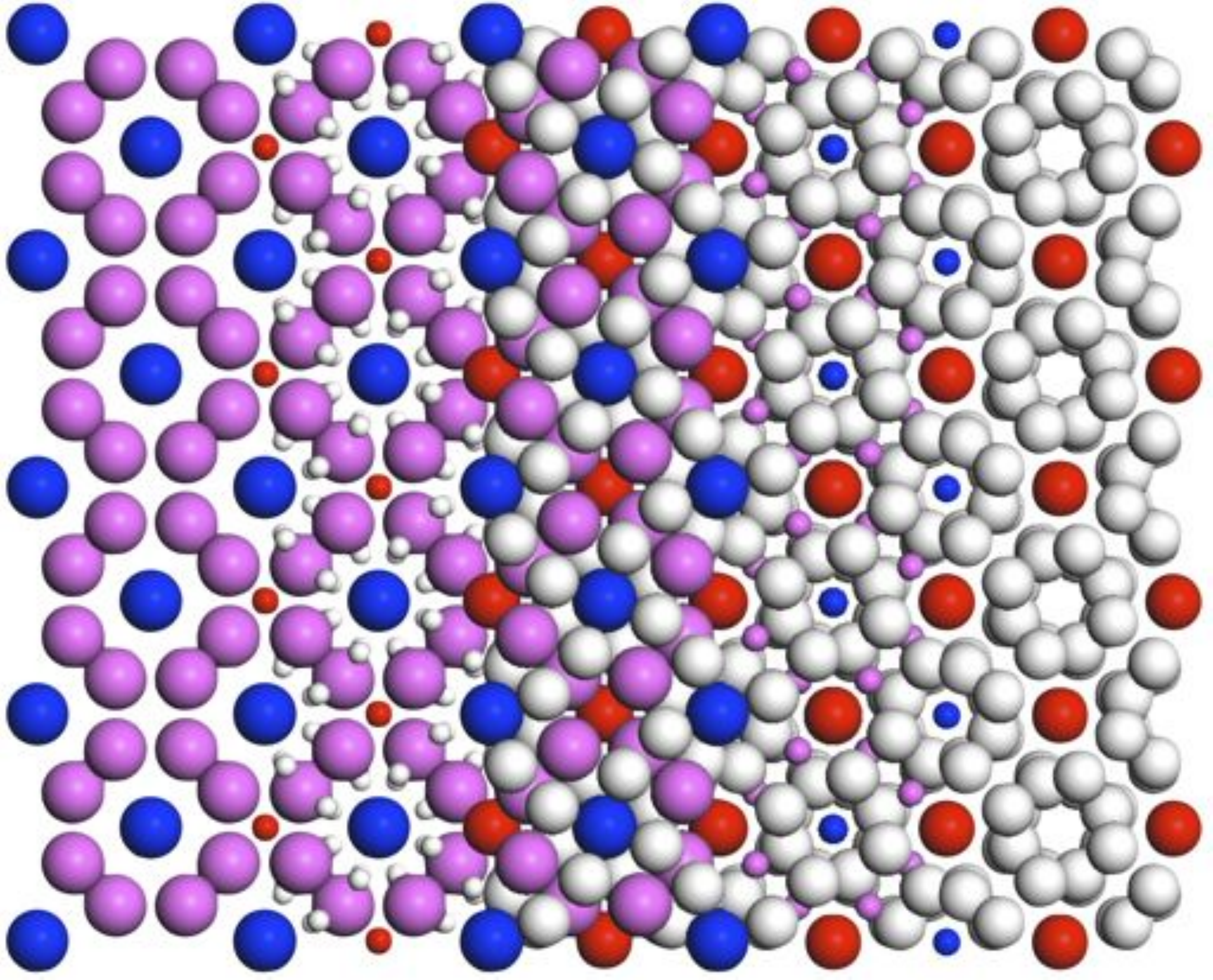}
\caption[]{A representation of the Ba-IV incommensurate host-guest
  structure is shown on the left, with the host atoms in purple and
  the guest atoms in blue.  The Ba-IV structure is also found in
  compressed Sr, Sc, As, Sb and Bi, and we predict it to be stable in
  aluminium in the range 3.2--8.8 TPa.  A representation of the Rb-IV
  incommensurate host-guest structure is shown on the right with the
  guest atoms in red and the host atoms in white.  The Rb-IV structure
  is found in Rb, K and Na at high pressures.  Both structures consist
  of positively charged ions and negatively charged electron blobs
  located within interstitial regions.  The Ba-IV and Rb-IV structures
  show a remarkable duality.  The electron blobs in the Ba-IV
  structure occupy the atomic positions of the Rb-IV structure, while
  in the Rb-IV structure the electron blobs occupy the atomic
  positions of the Ba-IV structure \cite{Pickard_2010_aluminium}.  The
  figure shows a view along the axis of the guest chains.  As we scan
  the picture from left to right the structure changes from Ba-IV to
  Rb-IV.}
\label{fig:al_host-guest.pdf}    
\end{figure}

Aluminium is used as a standard material in shock wave experiments,
for which purpose an accurate equation of state must be available.
Aluminium adopts the fcc structure under ambient conditions and
transforms to hcp at 0.217 terapascals (TPa) \cite{AkahamaNKK06}, and
a further transition to a body-centred-cubic (bcc) structure has been
predicted at 0.38 TPa using DFT methods \cite{TambeBM08}.  Our
searches have identified a transformation from bcc to the Ba-IV
non-close-packed incommensurate host-guest structure at 3.2 TPa and a
further transition to a simple hexagonal structure at 8.8 TPa
\cite{Pickard_2010_aluminium}.  The non-close-packed structures have
smaller volumes than bcc and their occurrence significantly alters the
high-pressure equation of state.  An important feature of our searches
was that we studied cells containing 2, 4 and 8--21 atoms.  Such a
systematic search can yield interesting results and we found
commensurate analogues of the host-guest structures in cells of 11,
16, and 21 atoms.  The physics behind the occurrence of
non-close-packed structures in highly compressed aluminium is similar
to that described above for lithium at much lower pressures.  The
simple hexagonal structure consists of alternate layers of aluminium
ions and electrons.  There are two ``blobs'' of electronic charge for
every ion and, considering the aluminium ions as the cations and the
electron blobs as the anions, the structure is that of magnesium
diboride (MgB$_2$), which is well known in ionic compounds of AB$_2$
stoichiometry.  We described the stability of the different structures
under pressure using empirical inter-atomic potentials to describe the
aluminium ions and electron blobs.  The potential parameters were
tuned to stabilise the host-guest structure, and it then gave the bcc
structure at lower pressures and the simple hexagonal structure at
higher pressures.  We also found a duality between the Ba-IV structure
and the other incommensurate host-guest structure found in the
elements, the Rb-IV structure, as explained in figure
\ref{fig:al_host-guest.pdf}.

\subsection{{\bf Iron:}}

\begin{figure}[ht!]
\centering
\includegraphics[width=0.75\textwidth]{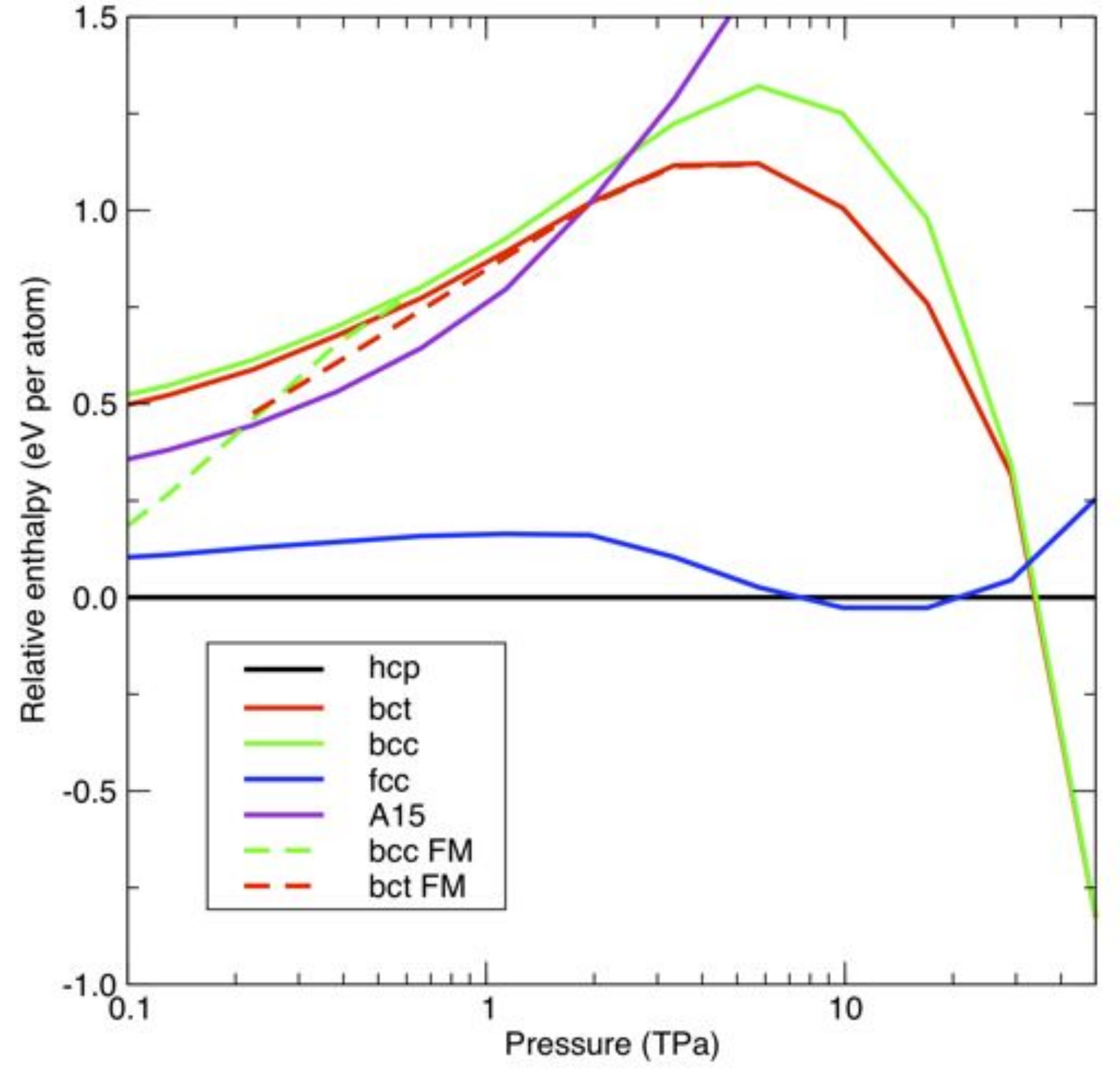}
\caption[]{Variation with pressure of the enthalpies of various phases
  of iron with respect to the hcp phase. The dashed lines indicate
  ferromagnetic (FM) phases and the solid lines indicate non-magnetic
  phases.}
\label{fig:Iron_enthalpy}    
\end{figure}

The Earth's core is largely composed of iron.  Other planets,
including many of the recently-discovered extrasolar planets (or
exoplanets), are expected to possess iron-rich cores.  Pressures
similar to those at the centre of the Earth have been achieved in
static diamond anvil cell experiments, but the multi-terapascal (TPa)
pressures expected at the centres of more massive planets can
currently be achieved only in shock-wave experiments, which give very
limited structural information.  Indeed, there are no materials whose
structures have been determined experimentally at pressures of 1 TPa
or more.  At low pressures the electronic configuration of the iron
atoms can be described as $3d^{6}4s^2$, but the more extended $4s$
orbitals are pushed up in energy with respect to the $3d$ orbitals
under compression and the $4s$ charge slowly drains into the $3d$
orbitals, leading to a $3d^{8}4s^0$ configuration at multi-TPa
pressures.  AIRSS showed that only the standard close-packed phases
are energetically competitive at multi-TPa pressures
\cite{Pickard_2009_iron}, see figure \ref{fig:Iron_enthalpy}.  The bcc
structure is stabilised at low pressures by its ferromagnetic spin
ordering, but it transforms to a hcp structure at pressures well below
100 GPa.  We found a transition from hcp to fcc and back to hcp at TPa
pressures (see figure \ref{fig:Iron_enthalpy}), although these
structures have similar enthalpies in the range 5--30 TPa.  The most
outstanding result was our prediction that the bcc phase, and a small
bct distortion of it, become much more stable than hcp and fcc at
extremely high pressures \cite{Pickard_2009_iron}.  The reason for
this is that the density of bct/bcc is about 0.6 \% higher than hcp at
the phase transition, which amounts to a very large enthalpy gain at
pressures of around 30 TPa.  We also studied harmonic phonon modes and
the effects of electronic excitations at finite temperatures, but the
overall effect on the relative stabilities of the phases is not large
\cite{Pickard_2009_iron}.

\subsection{{\bf Defects in silicon:}}

We have used AIRSS to study defect complexes in Si consisting of
combinations of H, N, and O impurity atoms and Si self-interstitials
and vacancies \cite{Morris_2008_H_in_Si,Morris_2009_HNO_in_Si}.  Most
of the searches were performed with 32-atom supercells, although we
used larger cells for a few searches.  We embedded the most
interesting defects in larger cells and relaxed them with a higher
energy cutoff and better k-point sampling.  We found almost all of the
previously-known point defects containing these impurity atoms, and we
also found a number of new lowest-energy defects for some
stoichiometries, such as $\{I,{\rm H}\}$ (an interstitial Si atom and
an H impurity atom) \cite{Morris_2008_H_in_Si}, and $\{3{\rm O}\}$
(three interstitial O impurity atoms) \cite{Morris_2009_HNO_in_Si}.
It is possible to automate the search procedure so that one needs
specify only the host crystal, the impurity atoms to be included and
the size and location of the ``hole'' in the host compound into which
the impurity atoms are placed.  The number of different combinations
of impurity atoms need not be excessive.  For example, using three
different types of impurity atom and a total number of impurity atoms
of $\leq$4 requires searching over only 34 possible cell contents, and
using five different types of impurity atom and a total number of
impurity atoms of $\leq$4 requires searching over only 125.  We
estimate that if we were presented with the crystalline structure of a
new material containing up to, say, three atomic species and we took
into account three possible impurity species (H, N, and O, for
example), we could determine the important point defects and their
physical and electronic structures within a few weeks.  Of course we
could also have predicted the structure of the host material.

\section{Conclusions}
\label{sec:conclusions}

The different searching methods which have been used in conjunction
with DFT methods should be judged by the results obtained.  We believe
that the AIRSS results presented here are impressive and that they
make a strong case for the method.  Our approach is pragmatic, we
start from the most random method for generating structures that we
can think of and introduce biases based on chemical, experimental
and/or symmetry grounds.  The starting structures are then relaxed
while preserving the experimental and symmetry constraints.  Sometimes
we perform shaking and/or phonon calculations on the relaxed
structures to look for energy lowering distortions.

We like the simplicity of our approach as it has a rather limited
number of ``knobs'' to turn whose effects are simple to understand.
This makes it easier to decide which knobs to turn and how far to turn
them, which allows more time for searching.

We concentrate our computational efforts on relaxing a very wide
variety of initial structures, which means that our stopping criterion
of obtaining the same lowest-energy structure several times gives a
good chance of finding the global minimum of the PES.

Our searching strategy will work very well on the petascale computers
which are becoming available now and the exascale computers which will
be available in a few years time.  Such computing resources will be
able to generate enormous databases of structures which will be useful
for many purposes, such as fitting and testing empirical force fields,
determining structures from diffraction data and determining
structures using data mining \cite{Fischer_2006}.  The efficient
handling and analysis of the huge amounts of data produced by
structure searches will pose challenges for the electronic structure
community.

Searching for structures with first-principles electronic structure
methods has already made an impact in various branches of science and
we imagine that it will become an integral part of materials design
and discovery.  Indeed it is reasonable to suppose that it will become
important in all fields in which it is relevant to know the relative
positions of atoms.

\section{Acknowledgements}
\label{sec:acknowledgments}

This work has been supported by the Engineering and Physical Sciences
Research Council (EPSRC) of the UK\@. 

\appendix

\section{Summary of other computational searching
  methods \label{sec:other_searching_methods}}

Although this article only deals with the AIRSS approach in detail, it
is appropriate to mention other techniques which have been used to
predict structures described by empirical or first-principles
inter-atomic forces.  There are many excellent reviews which describe
structure prediction methods for clusters and solids
\cite{Johnston_2003,Wales_book_2003,Baletto_2005,Woodley_2008,Rossi_2009,schon_2010}.

Simulated Annealing (SA) is a Monte Carlo technique devised by
Kirkpatrick \textit{et al} \cite{Kirkpatrick_1983}. The name derives
from an analogy with annealing in metallurgy, in which heating and
cooling is used to remove defects from a metal.  In this method the
current approximate solution or state is replaced by a randomly chosen
nearby state.  The probability of accepting the new state is 1 if it
is lower in energy than the initial state, and $e^{-\Delta E/T}$ if it
is higher, where $\Delta E$ is the energy of the final state minus the
initial state.  If the temperature $T$ is chosen to be zero then only
states of lower energy than the current state are accessible and the
algorithm normally becomes trapped in a local minimum.  To avoid this,
the temperature $T$ is gradually reduced during the simulation and, if
the cooling is slow enough, the system will eventually find the lowest
energy state.

SA with $T>0$ allows the system to jump out of local minima.  However,
the basic algorithm is normally inefficient as it often gets stuck in
local minima and many variants of it have been devised and tested in
the quest for higher efficiency.  There is considerable freedom to
alter the proposed moves and the form of the acceptance probability,
and to use more complicated ``annealing schedules'' in which the
temperature is sometimes raised during the run.

SA requires only the energies of different configurations of the system,
energy derivatives (forces and stresses) are not required.  It is, however,
straightforward to calculate energy derivatives using empirical potentials
and, with a little more effort, within first-principles methods.  Energy
derivatives can be used to replace the Monte Carlo algorithm by classical
molecular dynamics (MD).

The most widespread use of energy derivatives in structure searching is to
relax a structure to the minimum of its basin of attraction.

Methods have also been devised which evolve ensembles of structures
rather than evolving a single structure.  The simplest such idea is to
run entirely separate searches with different starting points.
Ensemble SA methods have been developed in which an adaptive annealing
schedule is controlled by ensemble averages of thermodynamic
information \cite{Ruppeiner_1991}.  Another idea is to use parallel
runs at different temperatures, such as in the parallel tempering
algorithm which derives from the work of Swendsen and Wang
\cite{Swendsen_1986,Earl_2005}.  The particle swarm method was
inspired by the collective behaviour of a flock of birds
\cite{Shi_1998}. In this MD-based method each member of the ensemble
or swarm is accelerated towards its own previous ``best solution'' and
towards the swarm's previous ``best solution''.

Locating the global minimum is difficult because the energy surface
contains many basins which may be separated by high barriers.  One
approach is to transform the energy surface to one which is easier to
search.  Perhaps the simplest such idea is to increase the range of
the inter-atomic potential \cite{Stillinger_1990} which has the effect
of removing many local minima \cite{Doye_1997}.  Such an unphysical
potential may of course have a significantly different global minimum.
Consider instead a transformed energy surface obtained by setting the
energy throughout each basin of attraction to the minimum energy of
the basin.  Obviously this transformation does not affect the relative
energies of the minima.  We now have to search the transformed energy
surface.  A simple Monte Carlo procedure known as ``basin hopping''
\cite{Li_1987,Wales_1999,Wales_book_2003} is to start at a random
position, relax to the basin minimum, propose a random move and relax
to the new minimum.  The move is accepted if the energy is lowered and
accepted with probability $e^{-\Delta E/T}$ if the energy is raised.
The simplest version of the algorithm has two parameters, the length
of the move which may be adjusted to give a reasonable acceptance
ratio, and the temperature $T$.  The minima hopping method
\cite{Goedecker_minima_hopping} is related to basin hopping.

EAs are optimisation techniques inspired by biological evolution,
involving concepts such as reproduction, mutation and recombination,
fitness and selection \cite{Back_1990}.  Genetic algorithms are a
subset of EAs in which a genetic representation of approximate
solutions (structures) is used, normally a bit array
\cite{Holland_1975}.  An ensemble or population of structures is
generated and each member is assigned a ``fitness'' which, for our
purposes, is its energy or enthalpy.  A fraction of the population is
selected for reproduction, with a bias towards the fittest, and they
are paired up for ``recombination'', the splicing together of the
parental genes.  A ``mutation'' step may also be performed.  The new
population is then subjected to selection and the whole process is
repeated.  When using EAs for structure searching it is standard to
relax structures to the minimum of their basin of attraction before
reproduction, so that the inheritance might be described as Lamarckian
rather than Darwinian.  EAs have been applied to many optimisation
problems, including LJ solids \cite{Oganov_glass_2008,Abraham_2008}
and clusters, and a review of the design and use of EAs for
determining the structures of atomic clusters described by empirical
potentials is given by Johnston \cite{Johnston_2003}.

The set of algorithms for predicting structures described above is of
course far from complete and interesting alternatives have been
pursued.  For example, crystalline network structures, such as
zeolites and carbon polymorphs, have been enumerated systematically
using graph theory \cite{Mackay_1991,Friedrichs_1999,Winkler_2001}.
Faken \textit{et al} \cite{Faken_1999} have sought high-dimensional
barrier-less pathways between local minima in the physical
three-dimensional space, and methods using quantum delocalisation have
also been investigated \cite{Amara_1993,Finnila_1994,Lee_2000}.
Metadynamics is a powerful sampling technique for reconstructing the
free-energy surface as a function of a set of collective variables,
and this method can be used to study phase transitions at finite
temperatures \cite{Laio_2002,Barducci_2008}.

Some of the strategies described above can clearly be combined, and
many additional refinements have been suggested.  There is often a
substantial overlap between the various different methods, and it can
be difficult to determine where one method ends and the next begins.
On reading the description of our AIRSS approach in Section
\ref{sec:random_structure_searching}, the reader will recognise
elements from the searching methods described in this appendix.

Almost all of the methods described above were first used in searching
for structures with empirical potentials, although they have since
been used with first-principles methods.  Jones and coworkers used
molecular dynamics simulated annealing with first-principles DFT to
study the structures of numerous clusters from the late 1980s
\cite{Hohl_1987,Hohl_1988}.  Deaven and Ho searched for cluster
geometries using an EA and a tight-binding model \cite{Deaven_1995},
and this work was important in bringing the possibilities of such
methods to the attention of the ``first-principles'' community.
Predicting crystal structures with first-principles methods is growing
in popularity.  Sch\"on, Jansen and coworkers have used Hartree-Fock
theory and DFT to search for stable structures and study the PES of
various crystals \cite{schon_2001,schon_2010}.  Zunger and coworkers
\cite{Hart_2005,Zhang_2010} and Oganov and coworkers
\cite{Oganov_2006,Oganov_boron_2009} have used EAs to search for
crystal structures with DFT methods.  Wang \textit{et al}
\cite{Wang_2010} have recently reported an application of a particle
swarm algorithm \cite{Shi_1998} to crystal structure prediction using
DFT methods.

\clearpage

\section{Details of new structures discussed in the text \label{sec:new_structures}}

\vspace{0.5cm}

\begin{tabular}{cclllllll}
Pressure       & Space group     & \multicolumn{3}{c}{Lattice parameters}           & \multicolumn{4}{c}{Atomic 
coordinates} \\
(GPa)          &                 & \multicolumn{3}{c}{(\AA, $^{\circ}$)}             & \multicolumn{4}{c}{(fractional)}       \\\hline
{\bf Boron}&&& \\
10             & $Cmcm$          & $a$=4.801      & $b$=8.710      & $c$=7.917      &B1 & 0.3176 & 0.5052 & 0.25
00           \\
               &                 & $\alpha$=90.00 & $\beta$=90.00  & $\gamma$=90.00 &B2 & 0.5000 & 0.5634 & 0.07
50           \\
               &                 &                &                &                &B3 & 0.2017 & 0.6685 & 0.13
82           \\
               &                 &                &                &                &B4 & 0.5000 & 0.7650 & 0.06
72           \\
               &                 &                &                &                &B5 & 0.3223 & 0.8328 & 0.25
00           \\
               &                 &                &                &                &   &        &        &     
             \\\\
50             & $C2/c$          & $a$=6.943      & $b$=4.669      & $c$=9.392      &B1 & 0.1858 & 0.0859 & 0.14
39           \\
               &                 & $\alpha$=90.00 & $\beta$=143.17 & $\gamma$=90.00 &B2 & 0.0054 & 0.4219 & 0.09
09           \\
               &                 &                &                &                &B3 & 0.3275 & 0.4141 & 0.18
82           \\
               &                 &                &                &                &B4 & 0.3368 & 0.5821 & 0.02
90           \\
               &                 &                &                &                &   &        &        &     
             \\\\
100            & $C2/m$          & $a$=5.164      & $b$=2.599      & $c$=7.568      &B1 & 0.0527 & 0.5000 & 0.13
07           \\
               &                 & $\alpha$=90.00 & $\beta$=143.36 & $\gamma$=90.00 &B2 & 0.7062 & 0.5000 & 0.54
70           \\
               &                 &                &                &                &B3 & 0.5192 & 0.5000 & 0.77
65           \\
               &                 &                &                &                &   &        &        &     
             \\\\ \hline
{\bf Nitrogen}&&& \\
10             & $Cmc2_1$        & $a$=5.336      & $b$=5.190      & $c$=7.930      &N1 & 0.5000 & 0.3502 & 0.00
08           \\
               &                 & $\alpha$=90.00 & $\beta$=90.00  & $\gamma$=90.00 &N2 & 0.5000 & 0.1559 & 0.11
36           \\
               &                 &                &                &                &N3 & 0.5000 & 0.2647 & 0.26
18           \\
               &                 &                &                &                &N4 & 0.5000 & 0.5184 & 0.23
41           \\
               &                 &                &                &                &N5 & 0.5000 & 0.5821 & 0.07
03           \\
               &                 &                &                &                &N6 & 0.5000 & 0.7155 & 0.34
07           \\
               &                 &                &                &                &   &        &        &     
             \\\\ \hline
{\bf LiH$_{16}$}&&& \\
100            & $I\bar4_2m$     & $a$=3.400      & $b$=3.400      & $c$=7.167      &Li1& 0.5000 & 0.5000 & 0.00
00           \\
               &                 & $\alpha$=90.00 & $\beta$=90.00  & $\gamma$=90.00 &H1 & 0.3780 & 0.2173 & 0.18
73           \\
               &                 &                &                &                &H2 & 0.7236 & 0.1164 & 0.07
73           \\\\
               
\end{tabular}

\end{document}